\documentclass[preprint]{aastex631}

\usepackage{amsmath}
\usepackage{mathrsfs}
\usepackage{boondox-calo}
\usepackage{xcolor}
\usepackage{CJK}
\usepackage{tabularx}
\usepackage{bm}
\DeclareRobustCommand{\uvec}[1]{{%
  \ifcsname uvec#1\endcsname
     \csname uvec#1\endcsname
   \else
    \bm{\hat{\mathbf{#1}}}%
   \fi
}}

\usepackage{gensymb}

\newcommand{\T}{\mathcal{T}}
\newcommand{\loss}{\text{loss}}
\newcommand\p{p}
\renewcommand{\deg}{\degree}
\newcommand\cross{\text{cross}}
\newcommand\U{U_\infty}
\newcommand\D{\mathcal{D}}
\newcommand\C{\mathcal{C}}
\newcommand\A{\mathcal{A}}
\newcommand\q{\mathcalo{q}}
\newcommand{\mean}[1]{\left\langle #1 \right\rangle}

\shorttitle{Scattering Dynamics}
\shortauthors{Huang et al.}

\graphicspath{{./}{figures/}}

\begin{document}

\begin{CJK*}{UTF8}{gbsn}
\title{Analytical Solutions for Planet-Scattering Small Bodies}

\author[0000-0003-1215-4130]{Yukun Huang (黄宇坤)}
\affiliation{Center for Computational Astrophysics, National Astronomical Observatory of Japan, 2-21-1 Osawa, Mitaka, Tokyo 181-8588, Japan}

\author[0000-0002-0283-2260]{Brett Gladman}
\affiliation{Dept. of Physics and Astronomy, University of British Columbia, 6224 Agricultural Road, Vancouver, BC V6T 1Z1, Canada}
\author[0000-0002-5486-7828]{Eiichiro Kokubo (小久保 英一郎)}
\affiliation{Center for Computational Astrophysics, National Astronomical Observatory of Japan, 2-21-1 Osawa, Mitaka, Tokyo 181-8588, Japan}

\correspondingauthor{Yukun Huang}
\email{yhuang.astro@gmail.com}

\begin{abstract}
Gravitational scattering of small bodies (planetesimals) by a planet remains a fundamental problem in celestial mechanics. It is traditionally modeled within the circular restricted three-body problem (CR3BP), where individual particle trajectories are obtained via numerical integrations. Here, we use {\"O}pik's close-encounter framework to study the random walk of the orbital energy $x$ for an ensemble of test particles on planet-crossing orbits. We show that the evolution of each particle's orbital elements $(a, e, i)$ is fully encapsulated by the 3D rotation of the relative velocity vector $\bm{U}_\infty$, whose magnitude remains constant. Consequently, the system can be reduced to two degrees of freedom. By averaging over all possible flyby geometries, we derive explicit expressions for the drift and diffusion coefficients of $x$. We then solve the resulting Fokker--Planck equation to obtain a closed-form solution for the time evolution of the particle distribution. A characteristic scattering timescale naturally emerges, scaling as $(P_{p}/M_{p}^{2})/500$, where $P_{p}$ is the planet's orbital period and $M_{p}$ its mass ratio to the central star. The typical ejection speed of small bodies by a planet is estimated to be $3 v_p M_{p}^{1/3}$, where $v_p$ is the planet's orbital speed. Our analytical solution constitutes a universal law applicable to both the Solar System and exoplanetary systems, providing a computationally efficient alternative to costly $N$-body simulations for studying the orbital distributions and ejection of planetesimals and planets (e.g., Kuiper Belt, Oort Cloud, debris disks, interstellar objects, and free-floating planets).

\end{abstract}

\keywords{Celestial Mechanics (211) --- Planetary dynamics (2173) --- Interstellar objects (52) --- Free floating planets (549) }

\section{Introduction} \label{sec:intro}
\end{CJK*}

The orbital evolution of small bodies (treated as massless particles) under the Newtonian gravity of a planet is a fundamental problem in celestial mechanics. While the problem is often modeled within the circular restricted three-body problem (CR3BP), assuming the planet's orbital eccentricity is negligible, it is well known that no general closed-form solutions exist to describe the resulting orbital evolution \citep{Szebehely.1968}. When the particle's and planet's orbits do not intersect, analytical insight can be obtained by expanding the perturbing function \citep[e.g.,][]{Murray.1999}. However, this approach is useful only when the particle's semimajor axis $a$ remains (nearly) constant, that is, during secular or resonant evolution.

In this work, we focus on scattering dynamics, in which a particle undergoes a random walk in semimajor axis due to repeated and stochastic close encounters with the planet. Strictly speaking, pure scattering dynamics, those entirely free of secular precession or temporary resonant sticking \citep{Duncan.1997, Gladman.2002}, do not exist. Nevertheless, we isolate the scattering process by concentrating on long-term variations in $a$ caused by random planet--particle encounters.

Planetary scattering occurs when a particle's orbit crosses the planet's and no mean-motion commensurability or von Zeipel--Lidov--Kozai cycle prevents arbitrarily close approaches\footnote{In the literature, a random-walk in semimajor axis for orbits with perihelion just beyond the planet's distance is also often called scattering. For example, in the Kuiper Belt, most scattering TNOs have $30 < q <  38$~au, beyond Neptune's orbit \citep[see][]{Gladman.2008}.}. Each encounter induces a stochastic change in the orbital elements, whose magnitude depends sensitively on the encounter geometry and the planetary mass. Because of this chaos, the particle's initial conditions are quickly forgotten, and the exact evolution is generally solved with a numerical integration. Even with the state-of-the-art algorithms, this is computationally expensive when applied to large ensembles of particles.

A related problem is diffusion in cometary and distant Kuiper Belt dynamics, where $a$ is much larger than the planet's orbital radius. \citet{Yabushita.1980} derived closed-form solutions for the evolution of comet ensembles; however, the energy diffusion coefficient $D$ was generally estimated by numerical experiments \citep{Fernandez.1981, Duncan.1987, Malyshkin.1999}. Recent surveys of trans-Neptunian objects (TNOs) have provided a larger sample to study the diffusion process \citep{Petit.2011, Bannister.2018, Bernardinelli.2022}. \citet{Batygin.202150u} and \citet{Belyakov.2026} showed that diffusion for non-crossing TNOs arises from overlapping chains of Neptune's mean-motion resonances, and \citet{Hadden.2024} used a mapping approach to derive analytical expressions for $D$ and the diffusion timescale of highly eccentric, non-crossing bodies.

However, neither series expansions nor mapping methods can handle crossing orbits, because when two objects can approach arbitrarily close to each other, the gravitational potential becomes singular. Hence, there is no general closed-form solution for the scattering timescale of crossing orbits. Here, we address this gap by providing an analytical description of the collective scattering of particle ensembles under {\"O}pik's close-encounter framework \citep{Opik.1951}.

This paper is organized as follows: In Section~\ref{sec:opik}, we introduce the {\"O}pik theory, which re-expresses a particle's heliocentric elements $(a,e,i)$ in terms of the planet-centric co-rotating coordinates $(U_\infty,\theta,\phi)$. In Section~\ref{sec:fokker-planck}, we derive expressions for the drift and diffusion coefficients and formulate the corresponding Fokker--Planck equation governing the energy evolution. In Section~\ref{sec:sol}, we present closed-form solutions to the Fokker--Planck equation, and validate our analytical results with numerical integrations. In addition, equations for the dynamical lifetime and ejection speed are also derived and verified. In Section~\ref{sec:applications}, we apply our analytical results to solar system giant planets and typical exoplanets.  Finally, in Section~\ref{sec:discussion}, we discuss more broadly future applications of our analytical framework. \\

\section{{\"O}pik's close-encounter theory}\label{sec:opik}

\subsection{3D Geometry in the Planet-centered Co-rotating Frame}\label{subsec:3d-geometry}

\begin{figure}[htb!]
  \centering
  \includegraphics[width=1.0\columnwidth]{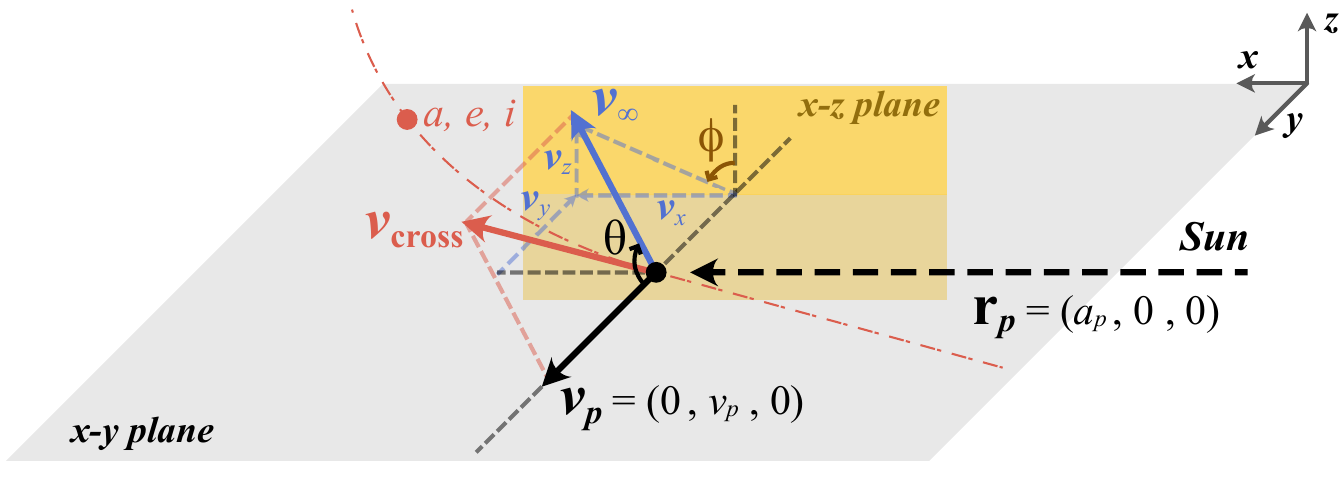}
  \caption{Schematic illustration of the 3D scattering geometry in a planet-centered co-rotating frame. The $z$-axis ($\uvec{k}$) points out of the orbital plane, while $x$-axis ($\uvec{i}$) and $y$-axis ($\uvec{j}$) are aligned respectively with the instantaneous radial and transverse directions of the planet at the time of encounter. In this coordinate system, $\bm{r}_p$ lies entirely along the $x$-axis, and $\bm{v}_p$ (the planet's velocity, solid black arrow) lies along the $y$-axis. The small body's heliocentric velocity vector at the intersection $\bm{v_\cross}$ (solid red arrow) is the vector sum of the relative velocity vector $\bm{v}_\infty$ (solid blue arrow) and $\bm{v}_p$. The vector $\bm{U}_\infty = \bm{v}_\infty/v_p$ is the normalized relative velocity. The two angles $\theta$ (the polar angle from $\bm{v}_p$) and $\phi$ (the azimuth angle of $\bm{v}_\infty$ in the $x$-$z$ plane) fully specify the direction of the relative velocity vector.}
  \label{fig:3d-scattering}
\end{figure}

We begin with a three-dimensional (3D) geometric representation of a planetary encounter. Assuming a planet (with a mass $m_p$) orbits the Sun (with a mass $m_\odot \gg m_p$) on a circular orbit of radius $a_p$ at a constant orbital speed $v_p = \sqrt{\mathcal{G}m_{\odot}/a_p}$, where $\mathcal{G}$ is the gravitational constant, a co-rotating reference frame centered on the planet is defined as follows (also shown in Fig. \ref{fig:3d-scattering}):

\begin{enumerate}
  \item The $x$-axis ($\uvec{i}$) always points from the Sun toward the instantaneous position of the planet  $\bm{r}_p$ .
  \item The $y$-axis ($\uvec{j}$) aligns with the instantaneous velocity of the planet $\bm{v}_p$.
  \item The $z$-axis ($\uvec{k}$) is perpendicular to the planet's orbital plane, following the right-hand rule.
\end{enumerate}

Consider a small body encountering the planet at a relative velocity $\bm{U}_\infty = \bm{v}_\infty/v_p$ (normalized to the planet's orbital speed), which can be expressed in Cartesian coordinates as
\begin{equation}\label{eq:def-relative-velocity}
  \begin{aligned}
    \bm{U}_\infty = U_{x}\ \uvec{i} + U_{y}\ \uvec{j} + U_{z}\ \uvec{k}.
  \end{aligned}
\end{equation}
The same vector can also be expressed in spherical coordinates \citep{Opik.1976, Carusi.1990}
\begin{equation}\label{eq:def-relative-velocity-spherical}
  \begin{aligned}
    \bm{U}_\infty = U_\infty \sin{\theta}\sin{\phi}\ \uvec{i} + U_\infty \cos{\theta}\ \uvec{j} + U_\infty \sin{\theta}\cos{\phi}\ \uvec{k},
  \end{aligned}
\end{equation}
where the angle $\theta$ is the polar angle measured from the $y$-axis (the direction of the planet's velocity $\bm{v}_p$), while $\phi$ is the counter-clockwise azimuthal angle measured in the $x$-$z$ plane from the $+z$ direction, looking back from the positive $y$ axis. These two angles are defined as:

\begin{equation}\label{eq:def-theta-phi}
  \begin{aligned}
    \cos{\theta} &= U_y /\U, \\
    \tan{\phi} &= U_x / U_z.
  \end{aligned}
\end{equation}

The particle's heliocentric velocity vector $\bm{v}_\cross$ is thus the vector sum of $\bm{v}_\infty$ and $\bm{v}_p$, whose magnitudes are related by
\begin{equation}\label{eq:def-vcross}
  \begin{aligned}
    \left(\frac{v_\cross}{v_p}\right)^2 &= \U^2 + 1 + 2 \U \cos{\theta}.
  \end{aligned}
\end{equation}

Having set up definitions for the velocity vectors in the rotating frame, heliocentric orbital elements $(a, e, i)$ follow directly \citep{Carusi.1990}. The dimensionless semimajor axis $\A \equiv a/a_p$, the orbital inclination $i$, and the eccentricity $e$ are given by

\begin{equation}\label{eq:def-A}
    \A = \frac{1}{1 -2 U_y - \U^2} = \frac{1}{1 -2 \U \cos{\theta} - \U^2},
\end{equation}

\begin{equation}\label{eq:def-e}
  e = \sqrt{\U^4 + 4U_y^2 + U_x^2(1-\U-2U_y) + 4\U^2U_y},
\end{equation}

\begin{equation}\label{eq:def-i}
    \cos{i} = \frac{U_y + 1}{\sqrt{U_z^2 + (U_y + 1)^2}},
\end{equation}

and conversely,

\begin{equation}\label{eq:aei-to-uxyz}
  \begin{aligned}
    U_x &= \sqrt{2 - 1/\A - \A(1-e^2)}, \\
    U_y &= \sqrt{\A(1-e^2)}\cos{i} - 1, \\
    U_z &= \sqrt{\A(1-e^2)}\sin{i}.
  \end{aligned}
\end{equation}

Rigorously speaking, a realistic approaching velocity vector $\bm{U}_\infty$ should include $\pm$ sign in front of $U_x$ and $U_z$ in Eq.~\eqref{eq:aei-to-uxyz}. The four combinations of $U_x$ and $U_z$'s signs correspond to four different encounter geometries (see \citealt{Valsecchi.1999}'s table 1), which we do not consider in this work. Instead, we use Eq.~\eqref{eq:aei-to-uxyz} to compute $U_x$ and $U_z$, which always yields positive $U_x$ and $U_z$. Consequently, $\phi \in (0, 90^\circ)$ for both prograde and retrograde orbits (Eq.~\ref{eq:def-theta-phi}). This framework allows one to have a consistent definitions of $U_x$, $U_z$, and $\phi$ across multiple encounters, which does not affect the following analysis of the scattering dynamics. Note that $U_y$ and $U_z$ in Eq.~\eqref{eq:aei-to-uxyz} are only valid in the limit of small minimum orbital intersection distance \citep{Vokrouhlicky.2012, Valsecchi.2022}. Further details are given in Section~\ref{subsec:scenarios}.

One can verify that these definitions also satisfy the Tisserand relation 
\begin{equation}\label{eq:def-tisserand}
  \begin{aligned}
    \frac{1}{\A} + 2\sqrt{\A(1-e^2)}\cos{i} = 3 - \U^2 \equiv \T,
  \end{aligned}
\end{equation}
where $\T$ is the Tisserand parameter relative to the planet. In the CR3BP, the Tisserand parameter is an approximate form of the Jacobi constant. This is also to say that the magnitude of the relative velocity $\U  = \sqrt{3-\T}$ is also invariant throughout multiple encounters.

In the {\"O}pik framework, specifying $\bm{U}_\infty$ (or equivalently $\U, \theta, \phi$) immediately yields the particle's $(a,e,i)$ in the heliocentric inertial frame. Therefore, the original scattering problem in which the particle experiences $(a,e,i)$ evolutions with constant $\T$, is reduced to the evolution of the two angles $(\theta, \phi)$ with constant $\U$. Note that this formalism does not track the argument of perihelion ($\omega$) or the longitude of the ascending node ($\Omega$), because each depends on the planet's position at the encounter, which we do not keep. In practice, however, the primary focus is on
$(a, e, i)$, so this omission is not problematic.

\subsection{Different Scenarios}\label{subsec:scenarios}
We illustrate in this section that the value of $\T$ (and $\U$) naturally determines the degree of coupling between the planet and the small body, in which case the applicability of the {\"O}pik framework presented in Section~\ref{subsec:3d-geometry} is different.

\begin{figure}[htb!]
  \centering
  \includegraphics[width=1.0\columnwidth]{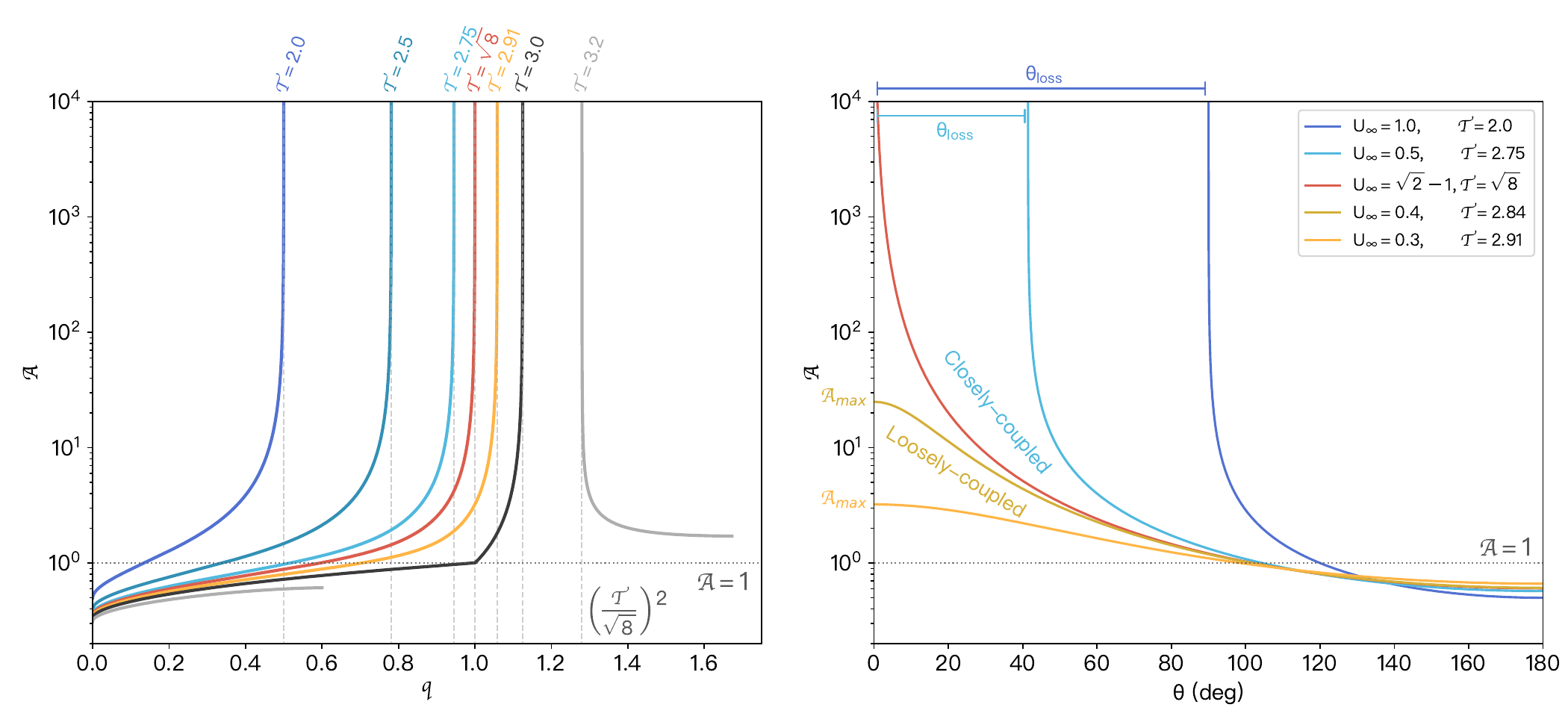}
  \caption{Left: Constant $\T$ curves on the dimensionless $\q$--$\A$ (perihelion distance--semimajor axis) space, assuming $i=0\deg$ (Eq.~\ref{eq:tisserand-a-q}). In the large-$\A$ limit, $\q$ asymptotically approaches the value $(\T/\sqrt{8})^2$ (vertical gray dashed lines, Eq.~\ref{eq:tisserand-a-infinity}). For $\T < \sqrt{8}$, $\lim_{\A \rightarrow \infty}{\q} < 1$ and the particle can never decouple from the planet in the co-planar case (blue). For $\T > \sqrt{8}$, $\lim_{\A \rightarrow \infty}{\q} > 1$ and spontaneous decoupling when the particle reaches large $\A$ is guaranteed (orange). The loosely-coupled case and the closely-coupled cases defined in text are separated by the critical value of $\T=\sqrt{8}$ (red), which has $\q \rightarrow 1$ as $\A \rightarrow \infty$. Eq.~\eqref{eq:tisserand-a-q} is partially undefined near the planet for $\T>3$ (grey).
  Right: $\A$ as a function of $\theta$ for various $\U$ (Eq.~\ref{eq:def-A}). The horizontal dashed line marks $\A = 1$. In a loosely-coupled case (golden curves), there is a maximum $\A$ the particle can reach in the planet-crossing state. For the boundary case $\U = \sqrt{2}-1$ (orange), $\theta \rightarrow 0\deg$ corresponds to $\A \rightarrow \infty$. In closely-coupled cases (blue), the object will escape the system once $\theta < \theta_\text{loss}$, which is defined in Eq.~\eqref{eq:def-thetaloss}. A larger $\U$ corresponds to a wider loss cone (blue line segments).}
  \label{fig:T-q-a-and-a-theta}
\end{figure}

One can re-write the definition of $\T$ (Eq.~\ref{eq:def-tisserand}) as a function of $\A$ and $\q$:
\begin{equation}\label{eq:tisserand-a-q}
    \T = \frac{1}{\A} + 2\sqrt{\q\left(2-\frac{\q}{\A}\right)} \cos{i},
\end{equation}
where the calligraphic lowercase $\q \equiv q/a_p$ is used to denote the dimensionless perihelion distance.
Since we are more interested in small bodies being scattered to highly-eccentric, large-$a$ orbits, numerical simulations suggest that such a evolution is often characterized by the quasi-conservation of the perihelion distance $q$ at large $a$. One can see that the conservation of perihelion is a natural consequence of the conservation of the Tisserand parameter when $\A \rightarrow \infty$:
\begin{equation}\label{eq:tisserand-a-infinity}
    \begin{aligned}
      \lim_{\A \rightarrow \infty}{\T} &= 2\sqrt{2\q} \cos{i},\\
      \lim_{\A \rightarrow \infty}{\q\cos^2{i}} &= \left(\frac{\T}{\sqrt{8}}\right)^2.
    \end{aligned}
  \end{equation}
Assuming $i = 0\deg$, we plot values of $\q$ as a function of $\A$ for a wide range of $\T$ in the left panel of Fig.~\ref{fig:T-q-a-and-a-theta}.

One can see from Eq.~\ref{eq:tisserand-a-infinity} and Fig.~\ref{fig:T-q-a-and-a-theta} that several critical values of $\T$ naturally divide the scattering problem into the following regimes:

\begin{enumerate}
\item \textbf{Diffusion regime}: $\U$ undefined ($\T > 3$).

One of the key assumptions in the {\"O}pik framework is that the small body's and the planet's orbits must intersect. The condition of $\T > 3$ breaks this criterion as small body is either completely outside the planet ($\q > 1$, see the gray curve in Fig.~\ref{fig:T-q-a-and-a-theta}) or completely inside (aphelion less than $a_p$). In this case, the close-encounter assumption fails, but the evolution can be studied in the diffusion approximation if the diffusion coefficient $D$ can be estimated.
This diffusion problem has been studied extensively by \citet{Batygin.202150u}, \citet{Hadden.2024}, and \citet{Belyakov.2026}.

\item \textbf{Loosely-coupled regime}: $\U < \sqrt{2} - 1$, for which $\sqrt{8} < \T < 3$.

For  $\U < \sqrt{2} - 1$, a near-planar orbit is coupled with the planet at lower $a$ while uncoupled at larger $a$, which is why we label it as the loosely-coupled case. The curve of constant $\T$ defines an outward escape route where encounters spontaneously raise $q$ above $a_p$ (see the yellow curve crosses $\q > 1$ in Fig.~\ref{fig:T-q-a-and-a-theta}). 
The {\"O}pik framework breaks up at a maximum semimajor axis $\A_\text{max}$, corresponding to where conservation of $\T$ requires a decoupling from the planetary orbit ($\q > 1$). 
One can solve for $\A_\text{max}$ by setting $\theta = 0\deg$ in Eq.~\eqref{eq:def-A}:
\begin{equation}\label{eq:def-amax}
  \A_\text{max} = \frac{1}{2-(\U+1)^2},\quad \text{if}\ \U < \sqrt{2} - 1 .
\end{equation}
We plot $\A$ as a function of $\theta$ in the right panel of Fig.~\ref{fig:T-q-a-and-a-theta}, with $\A_\text{max}$ labelled for two golden curves. The maximum semimajor axis, however, is only the upper limit for the {\"O}pik framework. In practice, a particle can still raise its $\A$ above $\A_\text{max}$ and escape the planetary system through weaker `energy kicks' at each perihelion passage.

  \item \textbf{Closely-coupled regime}: $\sqrt{2} - 1 < \U < \sqrt{2} + 1 $, for which $-\sqrt{8} < \T < \sqrt{8}$.

  When $\U > \sqrt{2} - 1$, the small body's heliocentric velocity after a flyby can exceed $v_\cross > \sqrt{2} v_p$, the escape velocity at $a = a_p$. The object can thus be ejected from the planetary system when the condition $\theta < \theta_\text{loss}$ is satisfied, which is often referred to as a \textit{loss cone} open on the unit sphere (see the right panel of Fig.~\ref{fig:T-q-a-and-a-theta}). The opening angle of the loss cone $\theta_\text{loss}$ is given by

  \begin{equation}\label{eq:def-thetaloss}
    \cos{\theta_\text{loss}} = \frac{1-\U^2}{2\U},\quad \text{if}\ \sqrt{2} -1 < \U < \sqrt{2} +1 .
  \end{equation}

  The critical value of $\T = \sqrt{8}$ yields $\q \rightarrow 1$ at $\A \rightarrow \infty$ (red curve in Fig.~\ref{fig:T-q-a-and-a-theta}). This serves as a separatrix between the loosely-coupled and the closely-coupled regimes.

  \item \textbf{Unbound regime}: $\U \geq \sqrt{2} + 1$, for which $\T \leq -\sqrt{8}$.

  This regime corresponds to $v_\cross \geq \sqrt{2} v_p$ for all $\theta$. All orbits are immediately unbound, with no repeated planetary encounters.
\end{enumerate}

In the general 3D case, even in the closely-coupled regime, a test particle can still be kicked to $\q > 1$ orbits, rendering the {\"O}pik model inapplicable. This can be seen by setting $U_x$ undefined in Eq.~\ref{eq:aei-to-uxyz}:

\begin{equation}\label{eq:Ux-undefined}
  \begin{aligned}
    2 - 1/\A - \A (1 - e^2) = \frac{\q^2}{\A} - 2\q + (2 - \frac{1}{\A}) < 0,
  \end{aligned}
\end{equation}
which is equivalent to $\q > 1$ in the case of $\A > 1$ \footnote{A similar constraint $Q/a_p < 1$ in the case of $\A < 1$ for undefined $U_x$ can also be derived, where $Q$ is the apocenter distance.}. Although $\theta$ and $\U$ can still be computed while $U_x$ is undefined, the physical picture of a particle encountering a planet is no longer valid when the two orbits are not crossing.

\begin{figure}[htb!]
  \centering
  \includegraphics[width=0.9\columnwidth]{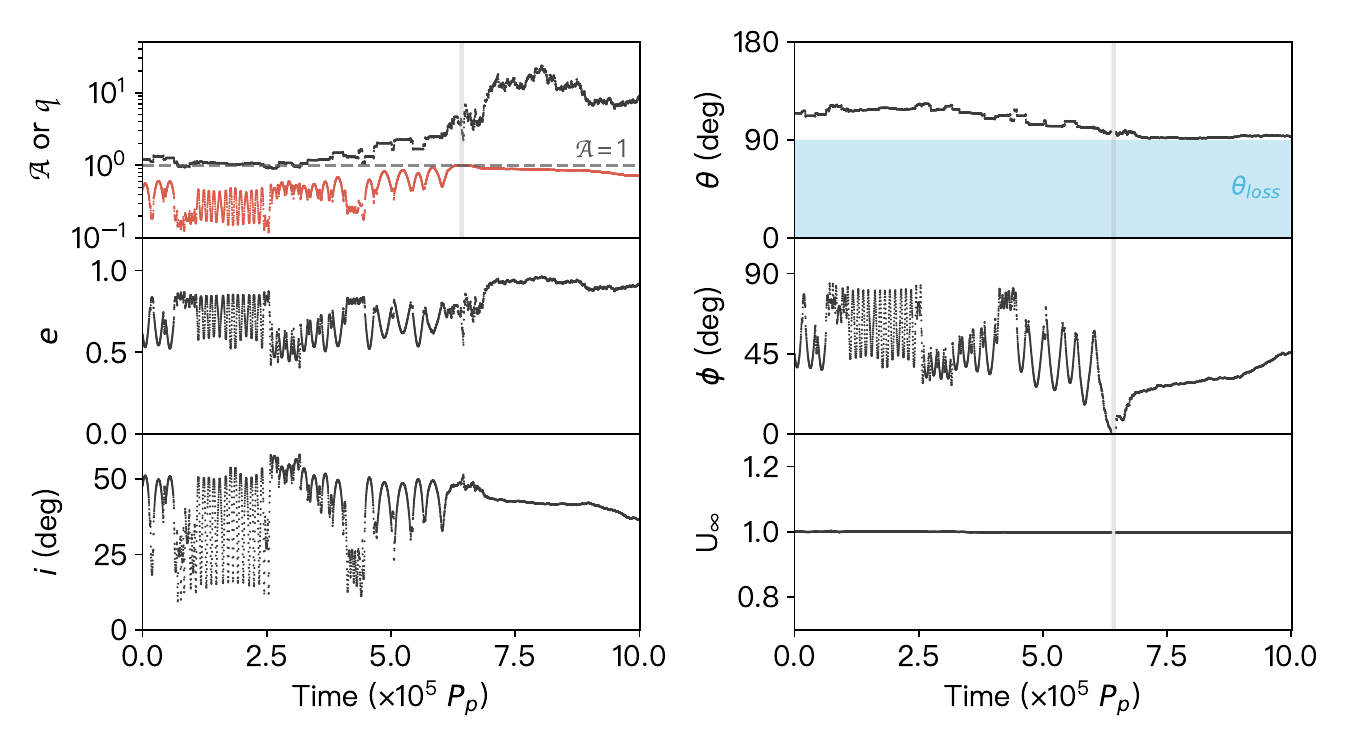}
  \caption{Evolution of a particle with $\U = 1.0$ ($\A_0$=1.2, $e_0$=0.61, $i_0$=47.8$^\circ$, closely-coupled case) in orbital elements ($\A$, $e$, $i$, left panels) and relative velocity vectors ($\theta$, $\phi$, $\U$, right panels) scattered by a $m_p = 1\times10^{-4} m_\odot$ planet on a $a_p = 1$~au circular orbit. The gray regions mark $\q > 1$ (red curves), corresponding to where the {\"O}pik framework fails. The blue region in the top-right panel represents the loss cone $\theta_\text{loss}$ (Eq.~\ref{eq:def-thetaloss}).}
  \label{fig:evolution_U1.0}
\end{figure}

\begin{figure}[htb!]
  \centering
  \includegraphics[width=0.9\columnwidth]{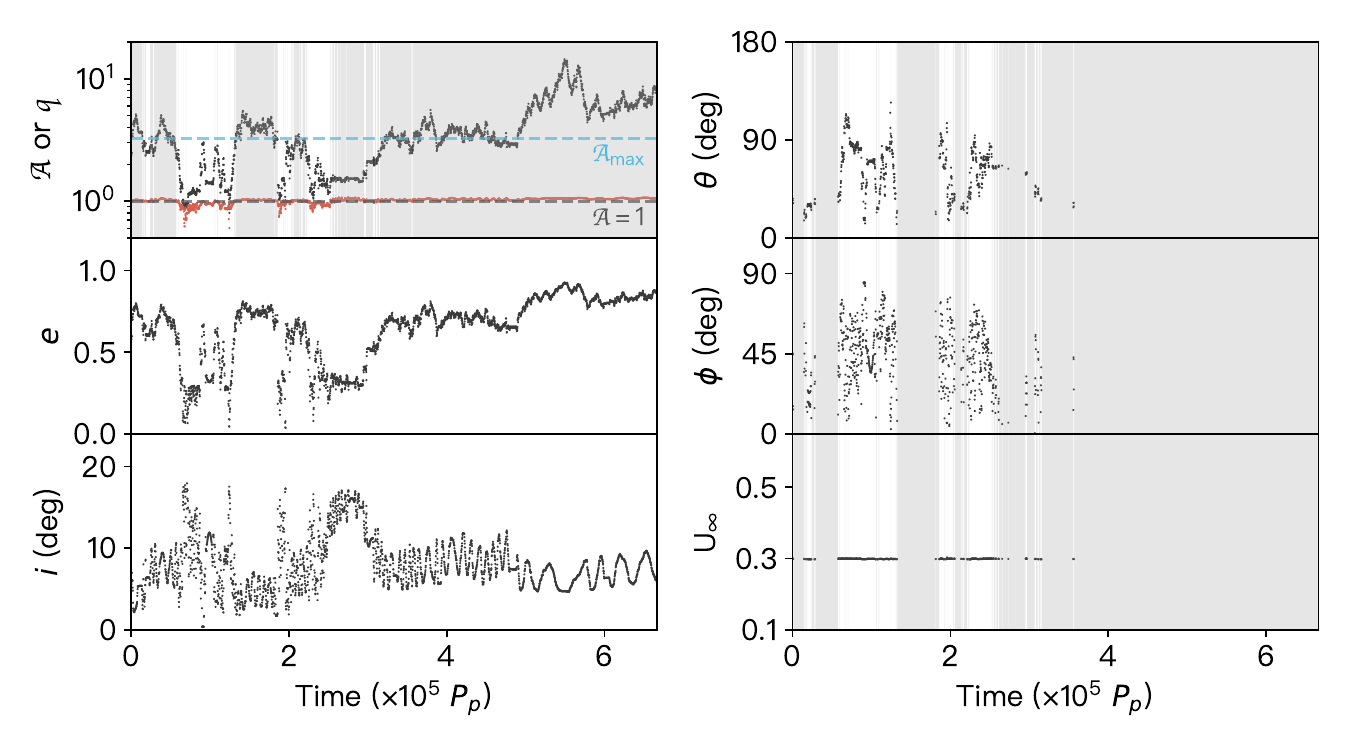}
  \caption{Same as Fig.~\ref{fig:evolution_U0.3}, but for a particle with $\U = 0.3$ ($\A_0$=2.4, $e_0$=0.59, $i_0$=5.5$^\circ$, loosely-coupled case). The blue dashed line denoted $\A_\text{max}$ (Eq.~\ref{eq:def-amax}) for this initial condition.}
  \label{fig:evolution_U0.3}
\end{figure}

In Figs.~\ref{fig:evolution_U1.0} and~\ref{fig:evolution_U0.3}, we present 
direct numerical CR3BP integrations of the evolution of two particles with $ \U = 1.0 $ (closely-coupled case) and $ \U = 0.3 $ (loosely-coupled case) in both orbital elements $(\A, e, i)$ and relative-velocity coordinates $(\theta, \phi, \U)$. As shown in Fig.~\ref{fig:evolution_U1.0}, the random walk in $\A$ is mapped onto a random walk in $\theta$ (Eq.~\ref{eq:def-A}). As $\A$ increases and approaches infinity, the corresponding value of $\theta$ asymptotically approaches the loss-cone boundary at $\theta_\loss = 90^\circ$, which corresponds to $\U = 1$ (Eq.~\ref{eq:def-thetaloss}). The time intervals shaded in gray indicate where the {\"O}pik theory becomes inapplicable, corresponding to $ \q > 1 $ in the top panels (red curves). For the closely-coupled particle in Fig.~\ref{fig:evolution_U1.0}, although temporary detachment from the planet ($\q > 1$) can still occur, it is confined to a relatively short interval when $ \phi $ is near $ 0^\circ $. This is in clear contrast to the loosely-coupled particle in Fig.~\ref{fig:evolution_U0.3}, where the {\"O}pik framework fails for most of the integration. 

Next, we discuss the evolution of $\phi$, defined in Eq.~\eqref{eq:def-theta-phi}. 
Strictly speaking, the true longitudinal angle describing the rotation of 
$\bm{v}_\infty$ around $\bm{v}_p$ in Fig.~\ref{fig:3d-scattering} should range 
from $0$ to $360^\circ$. However, it is more convenient to restrict $\phi$ to 
the first quadrant. The reason is as follows: for a particle on a crossing orbit 
with given $(a, e, i)$, the precession of $\omega$ causes the ascending and 
descending nodes to intersect the planet's orbit at four distinct locations\footnote{Although the longitude of the ascending node $\Omega$ also precesses, this \textit{nodal 
precession} does not affect the encounter geometry in CR3BP due to the longitudinal symmetry.}. 
These four \textit{nodal crossings} correspond to four possible encounter 
geometries, with $U_x$ and $U_z$ taking different $\pm$ signs according to 
Table~1 of \citet{Valsecchi.1999}. Nevertheless, all four longitudinal angles 
correspond to identical encounter probabilities, as we show later that the 
encounter probability is independent of $\omega$. Therefore, confining the 
longitudinal angle to $(0, 90^\circ)$ provides a consistent $\phi$ 
that characterizes the evolution of $(e, i)$. 

Additionally, both simulations demonstrate the conservation of $\U$ 
(bottom-right panels), which reflects the conservation of the Jacobi constant 
$\T$ in the CR3BP, independent of the {\"O}pik framework.

Although the $(\theta, \phi, \U)$ coordinates are strictly valid only for 
crossing orbits, they effectively reduce the scattering dynamics to two degrees 
of freedom. In this representation, $\theta$ characterizes variations in 
semimajor axis, whereas $\phi$ captures the coupled evolution of eccentricity 
and inclination. The random walk in $\theta$ serves as a clean signature of 
scattering, while the oscillations in $\phi$ are primarily driven by the 
planet's secular perturbations. This is evident from the $\phi$-oscillation 
period: it is short when $\A \approx 1$, but becomes significantly longer once 
the particle has been scattered to large semimajor axis (see the middle-right 
panel of Fig.~\ref{fig:evolution_U1.0}). Consequently, one cannot isolate an evolution of $ \phi $ due solely to scattering, since secular effects dominate. For this work, we therefore focus exclusively on the random walk of $\theta$, and treat $\phi$ statistically with reasonable assumptions.

\section{Fokker--Planck Equation}\label{sec:fokker-planck}

We have demonstrated that the orbital elements $(a, e, i)$ of a particle are uniquely determined by the relative velocity vector $\bm{U}_\infty$ in the planet-centered frame, which can be parameterized by $(\theta, \phi, \U)$. In this section, we derive the Fokker--Planck equation that describes the evolution of an ensemble of particles under planetary scattering. This requires one to average the effect of multiple encounters and compute the drift ($\mu$) and diffusion ($D$) coefficients, whose physical meanings are the mean and mean-square changes of the orbital energy per unit time.

\subsection{Averaging multiple encounters}\label{subsec:averaging}

\begin{figure}[htb!]
  \centering
  \includegraphics[width=0.5\columnwidth]{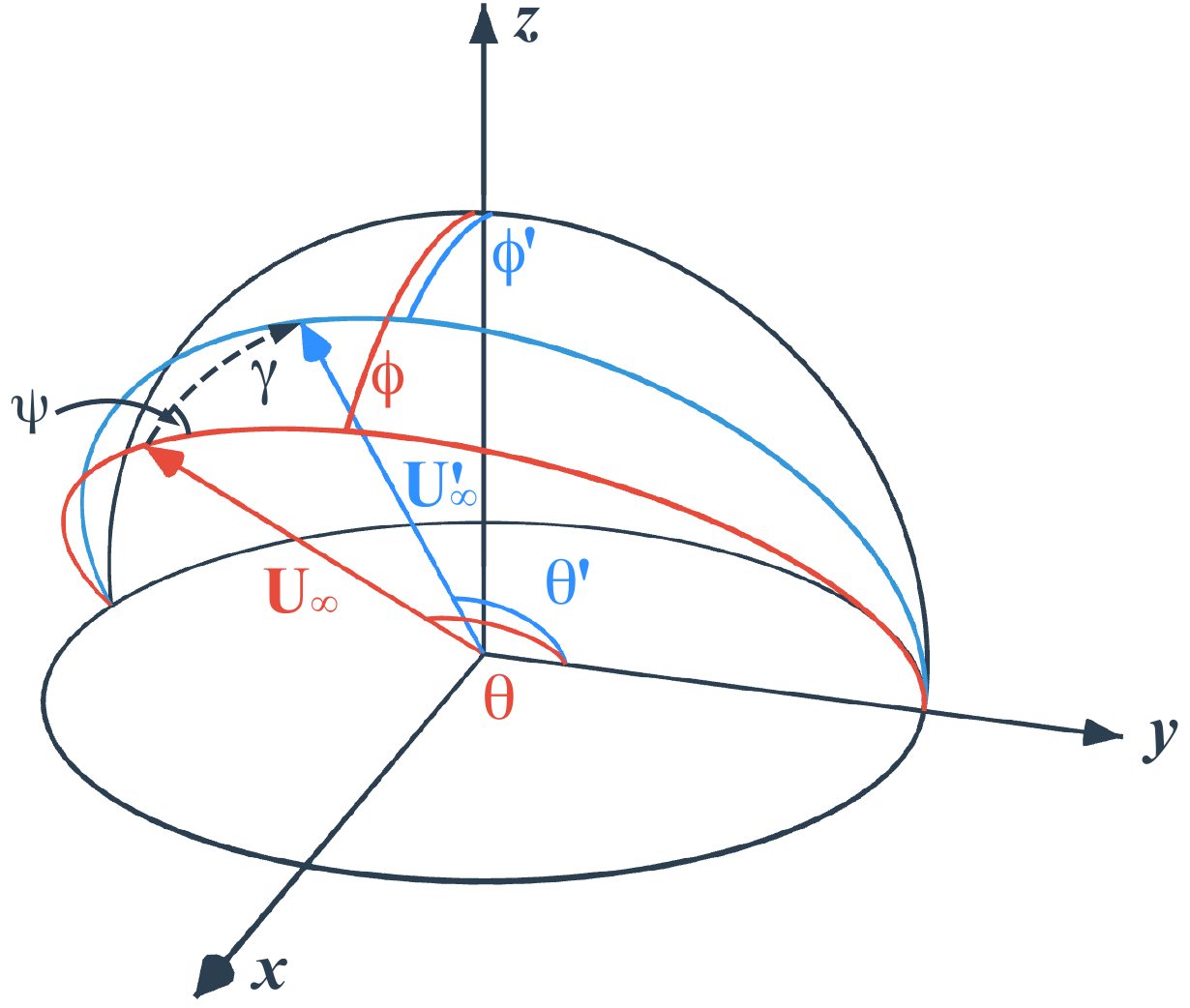}
  \caption{3D rotation of $\bm{U}_\infty$ (red) to $\bm{U}_\infty^\prime$ (blue) after a planetary encounter. The strength of the rotation is given by the angle $\gamma$ between the two vectors, whereas the direction of the rotation is oriented by the the counter-clockwise angle $\psi$.}
  \label{fig:3d_Uinfty_rotation}
\end{figure}

Consider a hyperbolic flyby with the impact parameter $b$ and the incoming velocity vector $\bm{U}_\infty$. 
After the encounter, $\bm{U}_\infty$ is rotated by $\gamma$ in some direction located by $\psi$, 
where $\psi$ is the counter-clockwise angle from the meridian containing $\bm{U}_\infty$ \citep{Opik.1976, Carusi.1990};  the 3D geometry of this rotation is shown in Fig.~\ref{fig:3d_Uinfty_rotation}.

The deflection angle $\gamma$ (the angle between the pre-flyby vector $\bm{U}_\infty$ and post-flyby vector $\bm{U}_\infty^\prime$) 
obeys
\begin{equation}\label{eq:def-gamma}
  \begin{aligned}
    \tan{(\gamma/2)} &= \frac{M_p}{B\U^2} \equiv C, \\
    \cos{\gamma} &= \frac{1 - C^2}{1 + C^2} \approx 1 - 2C^2 + O(C^4) \; ,
  \end{aligned}
\end{equation}
where $B = b/a_p$ and $M_p = m_p/m_\odot$ are the dimensionless impact parameter and planetary mass, respectively. We introduce another dimensionless parameter $C$ to denote the relative `strength' of a planetary encounter to change the particle's heliocentric orbital energy. By definition, the encounter strength increases ($\uparrow C$) for deeper ($\downarrow B$) and slower ($\downarrow \U$) encounters around a more massive ($\uparrow M_p$) perturber.

The mean-square strength of multiple flybys $\mean{C^2}$ is computed by averaging over all possible impact parameters $B$, a random variable following a $dN/dB \propto B$ distribution:
\begin{equation}\label{eq:mean-C}
  \begin{aligned}
    \mean{C^2} &= \left(\frac{M_p}{\U^2}\right)^2\cfrac{\int_{B_\text{min}}^{B_\text{max}} \frac{1}{B^2} B dB}{\int_{B_\text{min}}^{B_\text{max}} BdB},
  \end{aligned}
\end{equation}
where $B_\text{min}$ is the minimal impact parameter that would render the averaging approach invalid. In galactic dynamics, $B_\text{min} = B_{90}$ is often used, corresponding to the impact parameter deflecting the incoming velocity vector by $90^\circ$ \citep{Binney.2008};
$B_\text{max}$ is where the hyperbolic flyby assumption is no longer valid (e.g., the dimensionless Hill's sphere of the planet $R_H = r_H/a_p$):
\begin{equation}\label{eq:B_min_B_max}
\begin{aligned}
B_{\text{min}} &= B_{90} = M_p/U_\infty^2, \\
B_{\text{max}} &= R_H = \sqrt[3]{M_p/3},
\end{aligned}
\end{equation}
and the standard Coulomb ratio is
\begin{equation}\label{eq:Lambda}
  \begin{aligned}
    \Lambda \equiv B_{\text{max}}/B_{\text{min}} = R_H/B_\text{90} &=\U^2 / \sqrt[3]{3M_p^2}\ .
  \end{aligned}
\end{equation}

Since $\U^2$ is of order unity in the closely-coupled regime, $\Lambda \sim M_p^{-2/3} \gg 1$ for planets. In other words, the planetary Hill's sphere is significantly larger than the $90^\circ$ deflection impact parameter, which can be used to further simplify Eq.~\eqref{eq:mean-C} to
\begin{equation}\label{eq:mean-C-simp}
  \begin{aligned}
    \mean{C^2}
    &\approx  2\left(\frac{M_p}{\U^2 R_H}\right)^2 \ln\Lambda,
  \end{aligned}
\end{equation}
where
\begin{equation}\label{eq:ln_Lambda}
    \ln \Lambda = \ln\U^2 - \ln (M_p^2/3)/3,
\end{equation}
is the well-known Coulomb logarithm. 
This logarithm illustrates that close and distant encounters (between $B_\text{90}$ and $R_H$) both contribute meaningfully to the long-term scattering.

To further verify the applicable range of $B_{90}$, we compare it to the planetary radius,
\begin{equation}\label{eq:R_p}
    R_p = M_p^{1/3} R_\odot\left( \frac{\rho_p}{\rho_\odot}\right)^{-1/3},
\end{equation}
where $R_p$ and $R_\odot$ are the radii of the planet and the Sun in units of $a_p$. The ratio is
\begin{equation}\label{eq:B_90_to_R_p}
    B_{90}/R_p = \frac{M_p^{2/3}}{R_\odot \U^2}\left( \frac{\rho_p}{\rho_\odot}\right)^{-1/3}.
\end{equation}
Since both $\U^2$ and $(\rho_p/\rho_\odot)^{-1/3}$ are typically of order unity, Eq.~\eqref{eq:B_90_to_R_p} implies $B_{90}/R_p \sim M_p^{2/3}/R_\odot$. For Jupiter-mass, Neptune-mass, and Earth-mass planets orbiting a Sun-like star, we find $B_{90}/R_p \gtrsim 1$ for $a_p > 0.5$~au, $a_p > 4$~au, and $a_p > 40$~au, respectively. Thus, $B_{90}$ is easily attainable for giant planets, but is generally not reachable for terrestrial planets in the Solar System. This limitation is not significant, however, because $B_\text{min}$ only enters through the Coulomb logarithm, and changing $B_\text{min}$ by an order of magnitude alters Eq.~\eqref{eq:mean-C-simp} (and the resulting coefficients and timescales) by only a factor of $\sim$2.

One can see from Eq.~\eqref{eq:mean-C-simp} that $\mean{C^2} \sim M_p^{4/3}$, which is a small quantity in the case of planetary encounters. Thus, taking the leading terms of $\cos\gamma$ and $\cos^2\gamma$ from Eq.~\eqref{eq:def-gamma}, one obtains the following averages
\begin{equation}\label{eq:mean-cosgamma}
  \begin{aligned}
    \mean{\cos\gamma}
    &\approx 1 - 2\mean{C^2}, \\
    \mean{\cos^2\gamma}
    &\approx 1 -4\mean{C^2}.
  \end{aligned}
\end{equation}

Recalling the geometric relation between the pre-flyby vector $\U$ and the post-flyby vector $\U^\prime$ \citep{Carusi.1990}
\begin{equation}\label{eq:cos-theta-prime}
  \begin{aligned}
    \cos{\theta^\prime} &= \cos{\theta} \cos{\gamma} + \sin{\theta} \sin{\gamma} \cos{\psi}, \\
    \cos{\theta^\prime} - \cos\theta &= \cos{\theta} ( \cos{\gamma} -1) + \sin{\theta} \sin{\gamma} \cos{\psi},
  \end{aligned}
\end{equation}

and averaging $(\cos{\theta^\prime} -\cos\theta)$ and $(\cos{\theta^\prime} -\cos\theta)^2$ over $\gamma$ and $\psi$, one obtains
\begin{equation}\label{eq:mean-delta-costheta2-simp}
  \begin{aligned}
      \mean{\cos{\theta^\prime} -\cos\theta} &= -2\cos\theta\mean{C^2}, \\
    \mean{(\cos{\theta^\prime} -\cos\theta)^2} &= 2(1-\cos^2\theta)\mean{C^2},
  \end{aligned}
\end{equation}
in which the averaged quantities $\mean{\cos{\psi}} = 0$ and $\mean{\cos^2{\psi}} = 1/2$ are used, as $\psi$ is distributed isotropically over $(0,360\deg)$.

Next, we define
\begin{equation}\label{eq:def-x}
  \begin{aligned}
    x &\equiv \frac{\cos{\theta_\loss} - \cos{\theta}}{\cos{\theta_\loss} + 1}, \\
      &=\frac{1 - \U^2 - 2 \U \cos{\theta}}{1 + 2 \U  - \U^2},
  \end{aligned}
\end{equation}
where $x$ can be interpreted as a normalized position parameter defined in the range $(0, 1)$, with $x = 0$ corresponding to $\theta = \theta_\loss$ for which $\A \rightarrow \infty$, and $x = 1$ corresponding to $\theta = 180\deg$ and thus a particle with the smallest semimajor axis that still crosses the planetary orbit (see the right panel of Fig.~\ref{fig:T-q-a-and-a-theta}). It is important to note that throughout this paper, when $x$ appears as a subscript (e.g., in $U_x$), it denotes the $x$-component of a vector, whereas $x$ in its normal form denotes an independent variable.

Comparing Eqs.~\eqref{eq:def-x} and \eqref{eq:def-A}, one finds that
\begin{equation}\label{eq:x-a}
  \begin{aligned}
    x = \frac{1/\A}{1 + 2 \U  - \U^2}.
  \end{aligned}
\end{equation}

Thus, $x$ also has a physical relation to a particle's orbital energy. 
We define $x_{p} \equiv 1/(1 + 2 \U  - \U^2)$ 
as the value of $x$  when $\A = 1$, and write $x$ as
\begin{equation}\label{eq:x-a-simp}
  \begin{aligned}
    x =\frac{x_p}{\A},
  \end{aligned}
\end{equation}
where we recognize $x$ as a \textit{normalized} or \textit{dimensionless orbital energy} with respect to a target planet. This definition differs slightly the $x = 1/a$ notation often used in cometary studies, with an advantage being that this dimensionless $x$ is always normalized to $[0, 1]$, regardless of $a_p$ and $\U$.

We then use Eq.~\eqref{eq:mean-delta-costheta2-simp} to calculate the mean energy change $\mean{x^\prime -x}$ and the mean-squared energy change $\mean{(x^\prime -x)^2}$:
\begin{equation}\label{eq:mean-delta-x2}
  \begin{aligned}
    \mean{x^\prime -x} &= -\frac{\mean{\cos{\theta^\prime} -\cos\theta}}{\cos\theta_\loss+1} =2(x_{\perp}-x)\mean{C^2},\\
    \mean{(x^\prime -x)^2} &= \frac{\mean{(\cos{\theta^\prime} -\cos\theta)^2}}{(\cos\theta_\loss+1)^2} =2(1-x)(x-x_{\|})\mean{C^2},
  \end{aligned}
\end{equation}
where
\begin{equation}\label{eq:x_para}
    x_{\|} \equiv \frac{1-2\U-\U^2}{1+2\U-\U^2},\quad \quad x_{\perp} \equiv \frac{1-\U^2}{1+2\U-\U^2},
\end{equation}
 which are the respective values of $x$ when $\theta=0$ ($\bm{v}_\infty$ being parallel to $\bm{v}_p$, thus denoted with ${\|}$) and when $\theta=90\deg$ ($\bm{v}_\infty$ being perpendicular to $\bm{v}_p$, thus denoted with ${\perp}$).
 Notice that $x_{\|} < 0$ when $\U > \sqrt{2}-1$  (as the particle is ejected if ${\bf v}_\infty$ aligns with ${\bf v}_p$)
 and $x_{\|} \rightarrow -\infty$ when $\U \rightarrow \sqrt{2}+1$ (see Eq.~\ref{eq:def-x}). 
 Since $x \in (0, 1)$, $\mean{(x^\prime -x)^2}$ is only positive when $x_{\|} < 0$ ($\sqrt{2}-1 < \U < \sqrt{2}+1$), corresponding to the closely-coupled regime.

\subsection{Estimating the encounter probability}\label{subsec:encounter-probability}

To derive the drift coefficient $\mu(x)$ and the diffusion coefficient $D(x)$, we need to compute the time interval between two close encounters, and transform the mean (squared) energy changes from \textit{per encounter} to \textit{per unit time}. The intrinsic encounter probability (i.e., $B < R_H$ flybys) between a planet on a circular orbit and a particle is given by
\citep{Opik.1951, Wetherill.1967, Vokrouhlicky.2012} 

\begin{equation}\label{eq:Prob_B_RH}
    \text{Prob}(B< R_H) = R_H^2 \; \mathcal{F}(a, e, i) \; ,
\end{equation}
\noindent
in which the probability factor {\it per small body orbit} is:

\begin{equation}\label{eq:F_function}
  \mathcal{F}(a, e, i) \equiv \frac{\U}{\pi} \frac{1}{|\sin{i}|} {\sqrt{\frac{a/a_p}{(1-q/a_p)(Q/a_p- 1)}}} \; .
\end{equation}

Note that there are four singularities in $\mathcal{F}(a, e, i)$, corresponding to $i = 0$ or $180^\circ$, $q = a_p$, and $Q = a_p$ (where $Q$ is the dimensional aphelion distance), respectively. As explained in \citet{Vokrouhlicky.2012}, the pericenter and apocenter singularities stem from linearizing small displacements about the exact impact geometry; a more complete treatment removes these artifacts. In practice, these singularities simply reflect the substantial (but finite) increase in encounter probabilities when an orbit is nearly co-planar with the planet, or when its pericenter or apocenter distance lies close to the planet's orbital radius.

In the {\"O}pik framework, when $\phi = 0^\circ$, both $\bm{v}_{\infty}$ and $\bm{v}_{\rm cross}$ lie in the $(y,z)$-plane (Fig.~\ref{fig:3d-scattering}), implying that the incoming particle has essentially no radial heliocentric velocity component at the time of encounter. In this configuration, the particle must be extremely close to either perihelion or aphelion just before interacting with the planet, which greatly increases the likelihood of an encounter (Eq.~\ref{eq:F_function}). A similar enhancement occurs near $\phi = 90^\circ$, where the motion becomes nearly coplanar with the planet.

This geometric interpretation is fully consistent with the behavior of the {\"O}pik variables. As shown in Eq.~\eqref{eq:aei-to-uxyz}, the pericenter and apocenter singularities at $q = a_p$ and $Q = a_p$ correspond to $U_x = 0$, where $\phi = 0\deg$, while the inclination singularities at $i = 0^\circ$ and $180^\circ$ correspond to $U_z = 0$, where $\phi = 90\deg$, Thus, both boundaries yield substantially higher encounter probabilities than intermediate values of $\phi$. As a result, particles near $\phi=0\deg$ or $90\deg$ are scattered away far more rapidly than those with $\phi\sim45\deg$. Our numerical simulations confirm this: regardless of the initial conditions, the $\phi$-distribution of a particle ensemble rapidly relaxes to a steady state centered around $\phi \sim 45^\circ$ (see Appendix~\ref{app:steady-state-phi}). Therefore, to derive diffusion coefficients representative of the majority of particles, we hereafter adopt $\phi \sim 45^\circ$, or equivalently $U_x = U_z$ in Eq.~\eqref{eq:aei-to-uxyz}. This assumption will be applied throughout the derivation below.

\begin{figure}[htb!]
  \centering
  \includegraphics[width=0.7\columnwidth]{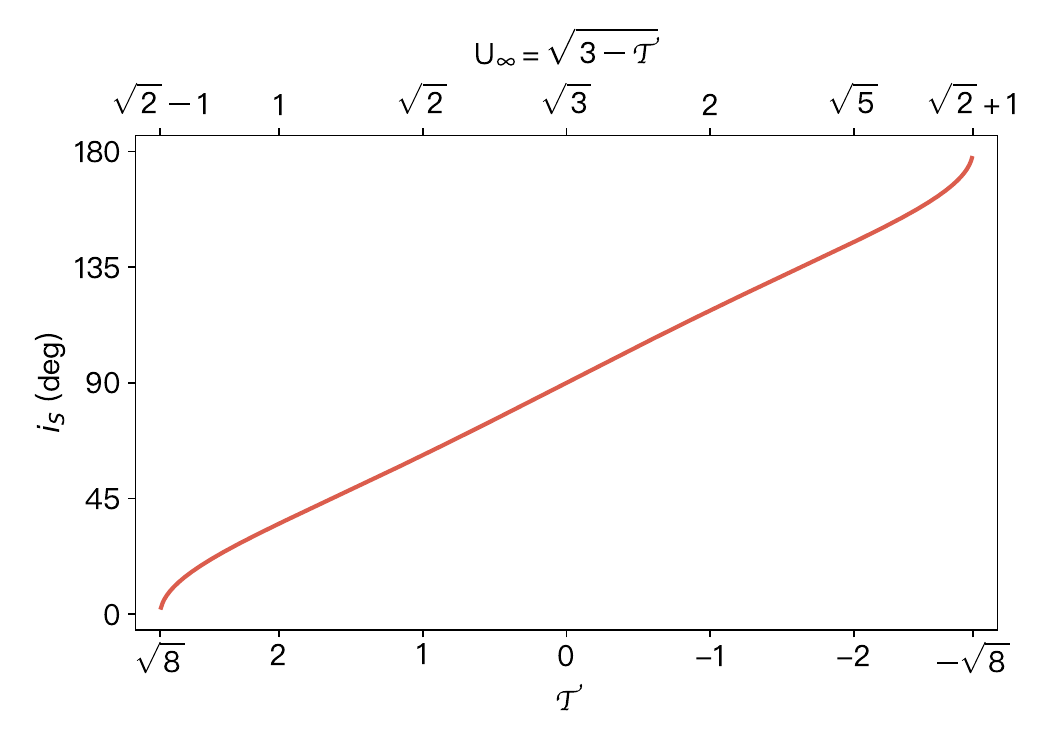}
  \caption{The typical scattering inclination $i_S$ as a function of $\T$ (or $\U$) when assuming $\phi = 45^\circ$. This represents the inclination to which most particles rapidly relax regardless of their initial inclinations.}
  \label{fig:iS}
\end{figure}

As we are mainly interested in the scattering dynamics in the large-$\A$ limit, we set $(Q/a_p - 1) \approx 2\A \gg 1$ in Eq.~\eqref{eq:F_function}, which yields
\begin{equation}\label{eq:F_function_2}
    \mathcal{F} \approx \frac{\U}{\pi} \frac{1}{|\sin i|} \frac{1}{\sqrt{2(1-\q)}}.
\end{equation}
Using $\q \approx \T^2/(8\cos^2 i)$ at large $\A$ (Eq.~\ref{eq:tisserand-a-infinity}), we obtain
\begin{equation}\label{eq:F_function_3}
    \mathcal{F} \approx \frac{\U}{\pi} \frac{2}{\sqrt{8\sin^2 i - \T^2 \tan^2 i}},
\end{equation}
which still retains an explicit dependence on inclination. Thus, an additional assumption regarding the `typical' inclination $i$ (or equivalently $\phi$) is required to evaluate the encounter probability in Eq.~\eqref{eq:F_function_3}. The typical scattering inclination $i_S$ in the large-$\A$ limit follows from adopting $\phi = 45^\circ$, as motivated above:
\begin{equation}\label{eq:typical-inclination}
    \sin i_{S} = \sqrt{\frac{8 - \T^2}{8 + \T^2}}, \qquad \text{for } -\sqrt{8} \leq \T \leq \sqrt{8},
\end{equation}
whose dependence on $\T$ (or $\U$) is shown in Fig.~\ref{fig:iS}.

With Eq.~\eqref{eq:typical-inclination},  $\mathcal{F}$ can be simplified to a function only depends on $\T$ ($\U$):
\begin{equation}\label{eq:F_function_approx}
  \mathcal{F} = \frac{2\sqrt{2}\,\U}{\pi} \frac{\sqrt{8 + \T^2}}{(8 - \T^2)},
\end{equation}
which can also be obtained by directly assuming $i = i_S$, $\phi = 45\deg$, and $\A \gg 1$ in \citet{Valsecchi.2022}'s equation~B.3.
Therefore, the average time interval between two close encounters $P_\text{enc}$ is the particle's period divided by the encounter probability

\begin{equation}\label{eq:P_enc}
  \begin{aligned}
    P_\text{enc} &= P_p \A^{3/2} / \ \text{Prob}(B< R_H), \\
    &= \frac{ P_p}{R_H^{2} \left(x/x_{p}\right)^{3/2} \mathcal{F}},
  \end{aligned}
\end{equation}
and the drift and diffusion coefficients (mean and mean-square changes per unit time) are given by
\begin{equation}\label{eq:mux_Dx}
  \begin{aligned}
    \mu(x) &= \mean{x^\prime -x} / P_\text{enc} =2\C(x_{\perp}-x)x^{3\over2}, \\
    D(x) &= \mean{(x^\prime -x)^2} / (2P_\text{enc}) =\C(1-x)(x-x_{\|})x^{3\over2},\\
  \end{aligned}
\end{equation}
where the rate parameter $\C$ is given by
\begin{equation}\label{eq:C}
  \begin{aligned}
    \C &= \frac{\mean{C^2} \; R_H^2 \mathcal{F}}{P_p \; x_{p}^{3/2}} = 
  \frac{2 \ln\Lambda}{P_p} \left(\frac{M_p}{\U^2}\right)^2 \frac{\mathcal{F}}{ x_{p}^{3/2}}.
  \end{aligned}
\end{equation}
It is worth noting that the $R_H^2$ factors in $P_\text{enc}$ (Eq.~\ref{eq:P_enc}) and $\mean{C^2}$ (Eq.~\ref{eq:mean-C-simp}) cancel each other out, and there is only a weak dependence on the Hill radius $R_H$ in the Coulomb logarithm of $\C$. The reason we write $\mu(x)$ and $D(x)$ is to obtain the Fokker--Planck equation, which describes the time evolution of an ensemble of particles scattered by a planet.

\subsection{Deriving the Fokker--Planck Equation}\label{subsec:fokker-planck}

Let $n(x, t)\mathrm{d}x$ be the number density of particles with normalized energy $x$ in the range $(x, x+\mathrm{d}x)$ at time $t$. Then, the Fokker--Planck equation that governs the evolution of $n(x, t)$ is given by
\begin{equation}\label{eq:Fokker_Planck_orig}
  \frac{\partial n(x,t)}{\partial t} + \frac{\partial S(x, t)}{\partial x} = 0,
\end{equation}
where
\begin{equation}\label{eq:def_Sxt}
  S(x,t) = \left[\mu(x) -\frac{\partial}{\partial x} D(x)\right] n(x,t)
\end{equation}
is the probability current. \citet{Hadden.2024} (hereafter HT24) proved that in a dynamical system whose Hamiltonian in action-angle variables $(J, j)$ is given by $\mathcal{H} = \mathcal{H}_0(J) + \epsilon \mathcal{H}_1(J, j, t)$, the following relation holds as long as the perturbation is small ($\epsilon \ll 1$):
\begin{equation}\label{eq:Hamiltonian_flow_relation}
  \mu = \frac{1}{J^\prime}\frac{\mathrm{d}}{\mathrm{d} x} \left(J^\prime D \right).
\end{equation}
In the restricted three-body problem, the action variable $J \propto \sqrt{a} \propto x^{-1/2}$ and its derivative $J^\prime \propto x^{-3/2}$, thus
\begin{equation}\label{eq:Hamiltonian_flow_relation2}
  \mu(x) x^{-3/2} = \frac{\mathrm{d}}{\mathrm{d} x} \left(x^{-3/2} D(x) \right).
\end{equation}
It is easy to verify that Eq.~\eqref{eq:Hamiltonian_flow_relation2} is equivalent to $2x_{\perp} = x_{\|}+1$, which is always true given their definitions in Eq.~\eqref{eq:x_para}.

To further simplify the Fokker--Planck equation, we define
\begin{equation}\label{eq:Dx}
  \begin{aligned}
    \D(x) &\equiv (1-x)(x-x_{\|}),\\
    D(x) &= \C\D(x)x^{3/2},
  \end{aligned}
\end{equation}
and plug it into Eq.~\eqref{eq:def_Sxt}:
\begin{equation}\label{eq:Sxt}
  \begin{aligned}
    S(x,t) &=  \left[\C \frac{\partial \D(x)}{\partial x} \left(x^{3/2} n(x,t)\right) -\frac{\partial}{\partial x} \left(\C\D(x)x^{3/2} n(x,t)\right)\right],\\
    &= -\C\D(x)\frac{\partial}{\partial x}\left(x^{3/2} n(x,t)\right).
  \end{aligned}
\end{equation}

The resulting Fokker--Planck equation reads
\begin{equation}\label{eq:Fokker_Planck_simp}
  \begin{aligned}
    \frac{\partial n(x,t)}{\partial t} = \C \frac{\partial}{\partial x}\left[\D(x)\frac{\partial}{\partial x}\left(x^{3/2} n(x,t)\right)\right],
  \end{aligned}
\end{equation}
which subjects to the following boundary conditions\footnote{We note a likely typo in HT24, where the boundary condition for their equation~(24) reads $n(0,t)=0$; based on private communication the intended condition is $\left.\left(x^{3/2}n(x,t)\right)\right|_{x=0}=0$. This does not affect the results reported in HT24.}
\begin{equation}\label{eq:boundary_conditions}
  \begin{aligned}
    \left.\left(x^{3/2} n(x,t)\right)\right|_{x=0} &= 0 \quad\text{(absorbing wall)}, \\
    \left.{\partial_x}\left(x^{3/2} n(x,t)\right)\right|_{x=1} &= 0 \quad\text{(reflecting wall)},
  \end{aligned}
\end{equation}
in which the absorbing wall at $x = 0$ states that a particle is removed when its energy $x$ becomes negative (hyperbolic ejection), whereas the reflecting wall $x = 1$ states that a particle cannot be removed from the problem when scattered inward. 
This essentially ignores physical collisions with the Sun when $q < r_\odot$, but should not significantly affect the result as long as the primary body's radius $r_\odot \ll a_p$.

\section{Analytical Solution and Validation}\label{sec:sol}

\subsection{Solution to the Fokker--Planck Equation}\label{subsec:fokker-planck-sol}

The Fokker--Planck equation we derived in Eq.~\eqref{eq:Fokker_Planck_simp} describes the time evolution of the probability density $n(x,t)$. Although a closed-form solution to the original equation does not exist, an approximate solution is obtained when one linearizes $\D(x)$ with a constant (remove its dependency on $x$). Given the simple form of $\D(x) = (1-x)(x-x_{\|})$ and that we are more interested in the large-$a$ scattering (i.e., $x \sim 0$), one way to linearize $\D(x)$ is simply setting $\D(x) = \D(0) = -x_{\|}$.

This results in the following Fokker--Planck equation that can be solved analytically
\begin{equation}\label{eq:Fokker_Planck_linearized}
  \begin{aligned}
    \frac{\partial n(x,t)}{\partial t} = \frac{1}{t_S} \frac{\partial^2\left(x^{3/2} n(x,t)\right)}{\partial x^2},
  \end{aligned}
\end{equation}
where $t_S = 1/(-x_{\|}\C)$ is the relaxation timescale, which we refer to as the ``scattering timescale'' in the following. Note that the constant $x_\parallel < 0$ if and only if $\sqrt{2} - 1< \U < \sqrt{2} + 1$ in Eq.~\eqref{eq:x_para}, thus the scattering timescale derived here is only physical in the closely-coupled regime ($-\sqrt{8} < \T < \sqrt{8}$).

Given an initial density $n(x, 0) = \delta{(x-x_0)}$, the solution to Eq.~\eqref{eq:Fokker_Planck_linearized} is given by (HT24)
\begin{equation}\label{eq:sol_Fokker_Planck_linearized}
  \begin{aligned}
    n(x,t) = \sum_{i=1}^{\infty} x^{-3/2} Y_i(x_0) Y_i(x) \exp\left[- {\lambda_i} \frac{t}{t_S}\right],
  \end{aligned}
\end{equation}
where the modes and the eigenvalues are defined by
\begin{equation}\label{eq:modes_and_eigenvalues}
    Y_i(x) = \frac{\sqrt{x} J_2(j_{1, i}\ x^{1/4})}{\sqrt{2} J_2(j_{1, i})}, \quad \quad
    \lambda_i = \frac{j_{1, i}^2}{16},
\end{equation}
in which $j_{1, i}$ is the $i$-th positive root of the first-order Bessel function of the first kind $J_1$, and $J_2$ is the second-order Bessel function of the first kind.
Using $\int_0^1 Y_i(x) x^{-3/2} \mathrm{d} x = \sqrt{2}/J_2(j_{1, i})$, the fraction of surviving particles as a function of time is obtained by integrating $n(x, t)$ from 0 to 1, which also
yields (HT24) an exponential decay:

\begin{equation}\label{eq:frac_survival}
  \begin{aligned}
    f_{\text{survive}}(t; x_0) = \sum_{i=1}^\infty \frac{\sqrt{x_0} J_2(j_{1, i}\ x_0^{1/4})}{\left|J_2(j_{1, i})\right|^2} 
       \exp\left[- {\lambda_i} \frac{t}{t_S}\right] \; .
  \end{aligned}
\end{equation}

We compare both the analytical solution (Eq.~\ref{eq:sol_Fokker_Planck_linearized}) and the fraction of survival particles (Eq.~\ref{eq:frac_survival}) with numerical integrations in Sec.~\ref{subsec:verification}.

\subsection{Scattering Timescale}\label{subsec:scattering_timescale}
The scattering timescale $t_S$ which appears in the exponential decay term of Eqs.~\eqref{eq:sol_Fokker_Planck_linearized} and ~\eqref{eq:frac_survival} is given by
\begin{equation}\label{eq:scattering_time}
  \begin{aligned}
    t_S = \frac{1}{(-x_{\|})\C} =  \frac{u}{\ln \Lambda} \left(\frac{P_p}{M_p^2}\right),
  \end{aligned}
\end{equation}
in which $u$ is a function that denotes the dependence of $t_S$ on the relative velocity $\U$, or equivalently the Tisserand parameter $\T$. One has
\begin{equation}\label{eq:uU}
  \begin{aligned}
    u(\U) &= \frac{ \U^4 x_p^{3/2}}{(-2 x_\parallel)\mathcal{F}}, \\
    & = \frac{\pi\U^3}{2}\left(2\U - 1 + \U^2 \right)^{-1/2}\left(\frac{2}{\U^4 - 6\U^2 + 17} - \frac{1}{8}\right)^{1/2},
  \end{aligned}
\end{equation}
and equivalently
\begin{equation}\label{eq:uT}
  u(\T) = \frac{\pi}{4\sqrt{2}} \sqrt{\frac{(3-\T)^3(8-\T^2)}{(2 - \T + 2\sqrt{3-\T})(8+\T^2)}},
\end{equation}
where the denominator also vanishes as $\T \to \sqrt{8}$, but a limit exists:
\begin{equation}\label{eq:uT_limit}
  \begin{aligned}
    \lim_{\T \to \sqrt{8}} u(\T) =  \lim_{\U \to \sqrt{2}-1} u(\U) = 0.0127.
  \end{aligned}
\end{equation}

Another factor in Eq.~\eqref{eq:scattering_time} that needs to be considered is the Coulomb logarithm $\ln \Lambda$ (Eq.~\ref{eq:ln_Lambda}), whose dependence on $\T$ and $M_p$ is significantly weaker:

\begin{equation}\label{eq:Ln_lambda_simp}
    \ln\Lambda = \ln{(3-\T)} - \ln(M_p^2/3)/3.
\end{equation}
The requirement that the Coulomb logarithm has to be positive gives the following constrain:
\begin{equation}\label{eq:max_Mp}
    M_p < \sqrt{3(3-\T)^3},
\end{equation}
which gives $M_p < 0.123$ at $\T = \sqrt{8}$. Since we assume $M_p \ll 1$, this constrain is naturally satisfied. The explicit expression of the scattering timescale $t_S$ is given by

\begin{equation}\label{eq:relaxation_time_general}
    t_S = \frac{\pi/(4\sqrt{2})}{\ln{(3-\T)} -\ln(M_p^2/3)/3} \sqrt{\frac{(3-\T)^3(8-\T^2)}{(2 - \T + 2\sqrt{3-\T})(8+\T^2)}}   \left(\frac{P_p}{M_p^2}\right),\quad \text{if}\ -\sqrt{8} < \T < \sqrt{8}.
\end{equation}

\begin{figure}[htb!]
  \centering
  \includegraphics[width=1.0\columnwidth]{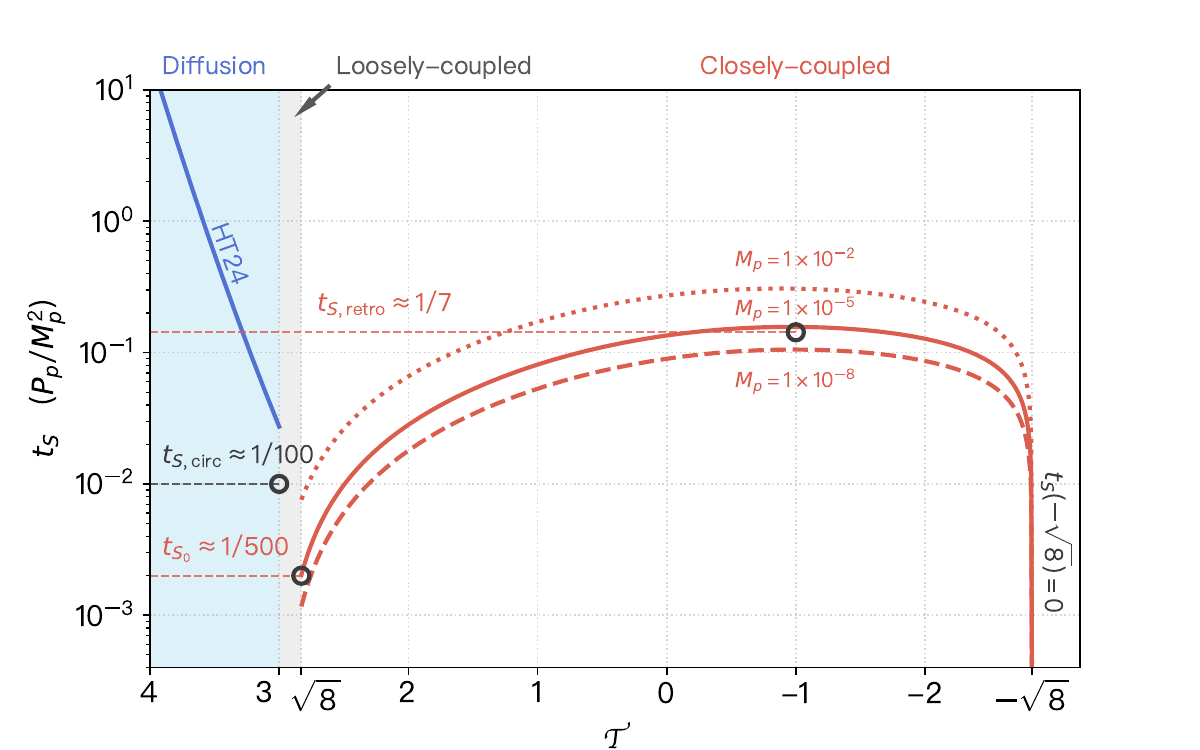}
  \caption{The scattering timescale $t_S$ (in the unit of $P_p/M_p^2$) as a function of $\T$ in different regimes. The red curves are computed using the full analytical expression (Eq.~\ref{eq:relaxation_time_general}) in the closely-coupled regime, with $M_p = 1\times10^{-2}$ (dotted), $1\times10^{-5}$ (solid), and $1\times10^{-8}$ (dashed), respectively. As $\T \rightarrow -\sqrt{8}$, $t_S$ shoots down to 0. The blue curve represents the best fit of the relaxation timescale in the diffusion regime, given by \citet{Hadden.2024} with a different approach. The narrow gap between the two regimes is the loosely-coupled regime, where neither models is valid. The minimum ($t_{S_0}$), circular ($t_{S, \text{circ}}$) and retrograde ($t_{S, \text{retro}}$) scattering timescales defined in text (Eqs.~\ref{eq:relaxation_time_sqrt8} and \ref{eq:relaxation_time_3}) are marked with black circles.}
  \label{fig:tS_vs_T}
\end{figure}

We plot $t_S$ (in the unit of $P_p/M_p^2$) as a function of $\T$ for $M_p = 1\times10^{-2}$ (dotted), $1\times10^{-5}$ (solid), and $1\times10^{-8}$ (dashed) in Fig.~\ref{fig:tS_vs_T}. One can see that regardless of $M_p$, the scattering timescale $t_S$ monotonically grows from $\T = \sqrt{8}$ to $-1$, corresponding to a near two-order-of-magnitude slowdown of the scattering process when a particle's typical inclination grows from $0\deg$ to $\approx$$120\deg$ (see Fig.~\ref{fig:iS}). Beyond $\T = -1$, the scattering timescale begins to drop, until it shoots down to 0 at $\T = -\sqrt{8}$ where all orbits were initially unbound.

The extremely weak dependence of $\ln \Lambda$ on $M_p$ can be seen in Eq.~\eqref{eq:Ln_lambda_simp}. As shown in Fig.~\ref{fig:tS_vs_T}, a three-order-of-magnitude difference in $M_p$ only leads to a factor of $\lesssim$2 difference in the Coulomb logarithm. Therefore, in terms of the planetary mass, the dominating factor is $M_p^{-2}$ in Eq.~\eqref{eq:relaxation_time_general}. For simplicity, we hereafter set $M_p = 1\times10^{-5}$ in $\ln \Lambda$ to obtain the following convenient forms of $t_S$ accurate to a factor of 2 (compared to Eq.~\ref{eq:relaxation_time_general}), suitable for all common planets with $M_p \sim (10^{-2}, 10^{-8})$, corresponding to 10 Jupiter mass and 0.1 Mercury mass, respectively, around a solar mass star:

\begin{equation}\label{eq:relaxation_time_sqrt8}
  \begin{aligned}
    t_{S_0} &= \left(\frac{P_p}{M_p^2}\right) \big/\ 500, \\
    t_{S, \text{retro}} &= \left(\frac{P_p}{M_p^2}\right) \big/\ 7,
  \end{aligned}
\end{equation}
where we call $t_{S_0}$ the \textit{minimum} scattering timescale, at $\T = \sqrt{8}$, and $t_{S, \text{retro}}$ is the \textit{retrograde} scattering timescale around $\T \sim -1$ (for which $i_S \sim 120\deg$).

For $-1 < \T < \sqrt{8}$, the slowdown of the scattering process is mainly contributed by a faster encounter velocity $\U$; whereas for $\T > \sqrt{8}$, where the particle's perihelion starts to decouple from the planet's orbit, the slowdown of the scattering is also inevitable as now the particle has to encounter the planet at a larger distance, even outside the planet's Hill sphere (distant encounters).

Although there is no closed-form expression for $t_S$ in the loosely-coupled regime (grey shaded region in Fig.~\ref{fig:tS_vs_T}), in the diffusion regime where orbits rigorously do not intersect (blue shaded region), one can compare the diffusion timescale (still referred to as $t_S$) derived by HT24 using a mapping approach with our scattering timescale in Fig.~\ref{fig:tS_vs_T}.  
By rescaling HT24's equation~(25) and replacing $q$ with $\T^2/8$ (assuming $i = 0^\circ$), one obtains\footnote{Note that in HT24, the diffusion timescale $T_D$ they use is $t_S/2$ in our notation.}:
\begin{equation}\label{eq:diffusion_timescale_Q}
    t_S = 6.4 \times 10^{-6}   \exp{[0.93\ \T^2]} \left(\frac{P_p}{M_p^2}\right),\quad \text{if}\ \T \gtrsim 3,
\end{equation}
which is plotted with the solid blue line in Fig.~\ref{fig:tS_vs_T}. The shortest diffusion timescale (corresponding to those near $\T = 3$) is roughly an order of magnitude longer than $t_{S_0}$\footnote{Note that in HT24's figure~6, the real diffusion timescale is even smaller than the best-fit power-law by a factor of 2, implying $t_S/t_{S_0} \sim 5$ at $\T = 3$.}. It is reassuring that, despite being derived from completely different methods, these two timescales agree with each other to a surprisingly high degree. Furthermore, they both constrain the scattering (diffusion) timescale in the narrow loosely-coupled regime ($\sqrt{8} < \T < 3$) to a factor of several $t_{S_0}$, consistent with previous numerical simulations (see \citealt{Huang.2023t}'s figure~3.12).

Therefore,  Fig.~\ref{fig:tS_vs_T} suggests a practical estimate for the scattering (diffusion) timescale $t_S$ at $\T = 3$, corresponding to initially circular and near (but not strictly) coplanar particles close to the planet's orbit:
\begin{equation}\label{eq:relaxation_time_3}
  t_{S, \text{circ}} \approx \left(\frac{P_p}{M_p^2}\right) \big/\ 100,
\end{equation}
and the scattering timescale is expected to smoothly grow from $t_{S_0}$ to $t_{S, \text{circ}}$ in the loosely-coupled regime, and then naturally transition to the timescale of the diffusion regime.

\subsection{Validation with Numerical Integrations}\label{subsec:verification}

\begin{figure}[htb!]
  \centering
  \includegraphics[width=1.0\columnwidth]{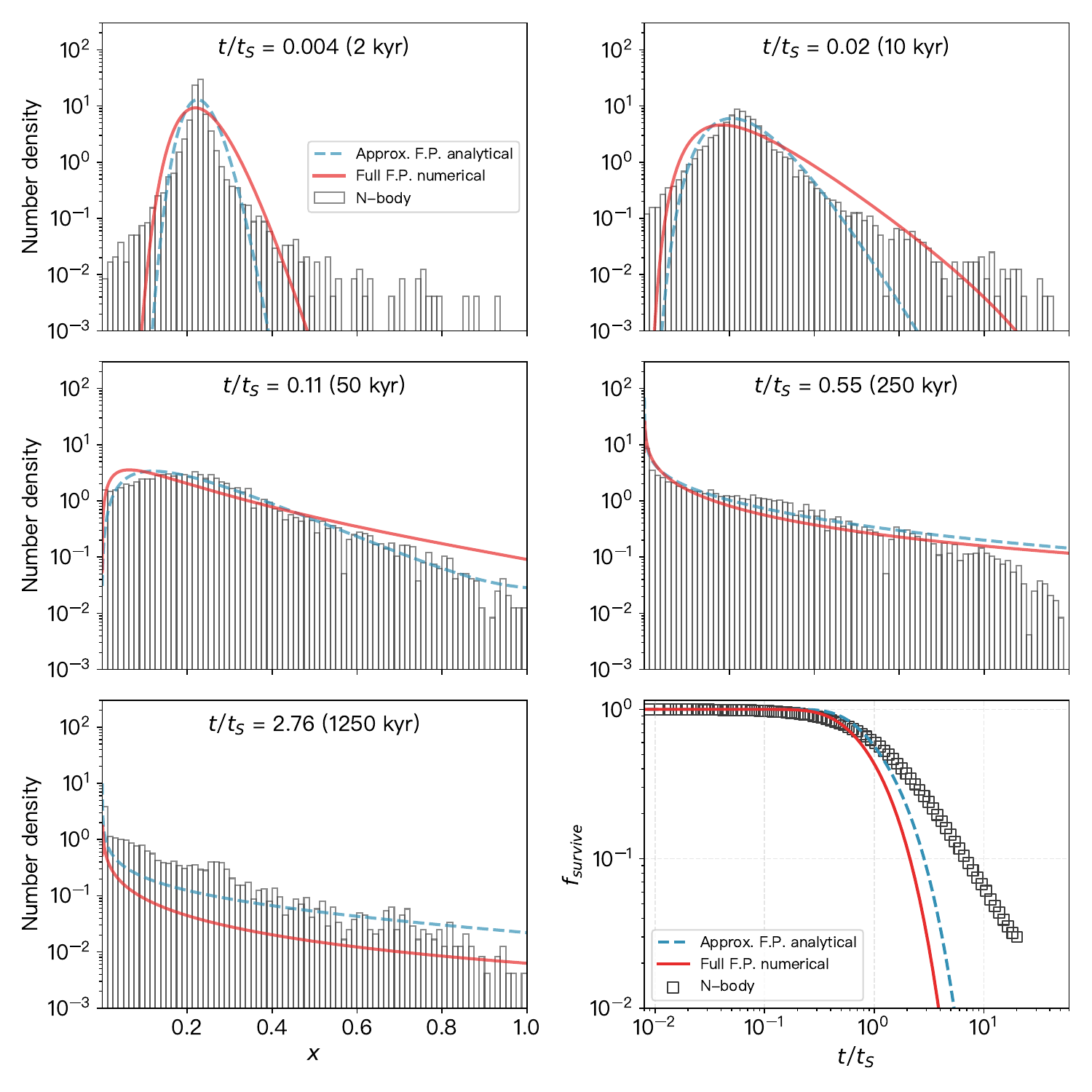}
  \caption{Comparison of the particle energy distribution ($x$) at various times predicted by the Fokker--Planck equation (solid and dashed curves) and by direct numerical integrations (bar histograms). The blue curves show the analytical solution of the linearized (approximate) Fokker--Planck equation (Eq.~\ref{eq:sol_Fokker_Planck_linearized}), while the red curves correspond to the numerical solution of the full Fokker--Planck equation (Eq.~\ref{eq:Fokker_Planck_simp}). Particles are initialized with $\A_0 = 2.49$, $e_0 = 0.634$, and $i_0 = 15.8\deg$, corresponding to $\U = 0.5$ or $\T = 2.75$. The planet has mass $M_p = 1\times10^{-4}$ at $a_p = 1$~au, and its scattering timescale $t_S = 450$~kyr is computed using Eq.~\eqref{eq:relaxation_time_general}. The final panel shows the surviving fraction $f_{\text{survive}}$ as a function of time, with the blue dashed curve given by Eq.~\ref{eq:frac_survival}.}
  \label{fig:hist_U05}
\end{figure}

\begin{figure}[htb!]
  \centering
  \includegraphics[width=1.0\columnwidth]{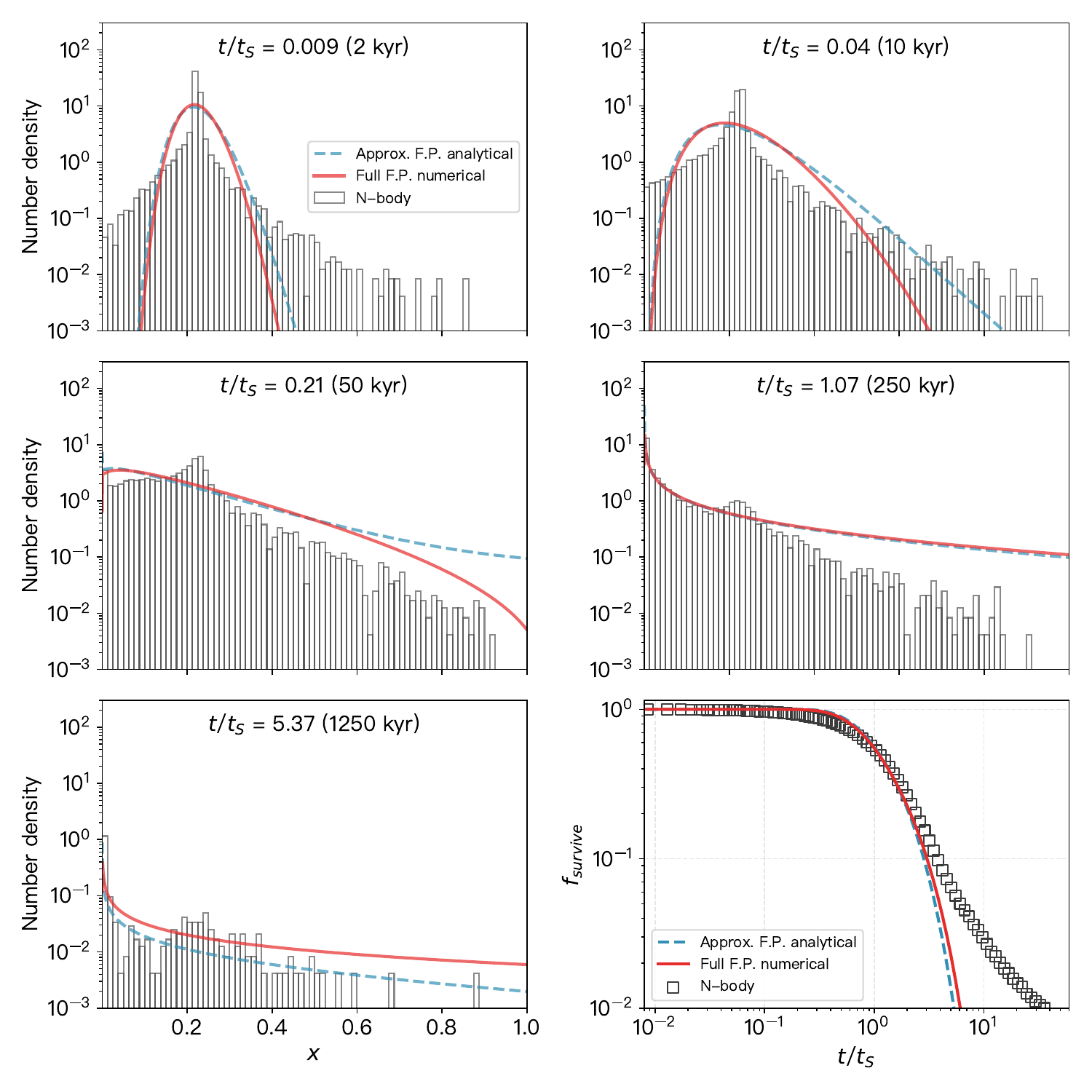}
  \caption{Same as Fig.~\ref{fig:hist_U05}, but for  $\A_0 = 4.47$, $e_0 = 0.871$, $i_0 = 126.2\deg$ particles, corresponding to $\U = 2$ or $\T = -1$. The planet has mass $M_p = 1 \times 10^{-3}$ instead, and its scattering timescale $t_S = 230$~kyr.}
  \label{fig:hist_U2}
\end{figure}

We now compare both the numerical solution of the full Fokker--Planck equation (Eq.~\ref{eq:Fokker_Planck_simp}) and the analytical solution of the approximated Fokker--Planck equation (Eq.~\ref{eq:sol_Fokker_Planck_linearized}) to numerical integrations in Fig.~\ref{fig:hist_U05} and Fig.~\ref{fig:hist_U2} (see figure captions for their respective initial conditions). The initial semimajor axes of all the test particles are carefully chosen so that they are not close to major mean motion resonances with the planet. Both integrations initially have 19,200 particles, and are performed in REBOUND \citep{Rein.2012} with the \texttt{Mecurius} \citep{Rein.2019} hybrid symplectic integrator that can handle close encounters. We do not consider particle's collisions with the planet or the Sun, which are also not modeled in the analytical framework.

Both Fig.~\ref{fig:hist_U05} and Fig.~\ref{fig:hist_U2} show reasonable agreement between the analytical predictions and the numerical integrations over a wide range of times spanning several orders of magnitude. In both cases, the Fokker--Planck equation under predicts the rapid initial spread in the energy distribution (see the first panels), which is expected given that we used an averaging approach to derive the diffusion coefficient. In practice, a particle can still be randomly scattered to a significantly different orbit (i.e., a large jump in $x$) even at the beginning of the integration.  

In the retrograde case, the distribution peak around the initial $x_0 \approx 0.22$ persists longer than the analytical theory predicts (Fig.~\ref{fig:hist_U2}, first four panels). Upon closer inspection, this is due to a portion of test particles being initialized inside the von Zeipel--Lidov--Kozai resonance, which also operates for retrograde orbits and phase protect them from having close encounters \citep{Huang.2018, Huang.2019}.

The Fokker--Planck equation also does not account for mean-motion resonances, yet resonance sticking (where particles are temporarily trapped in a planet's mean-motion resonance, prolonging their survival time, \citealt{Malyshkin.1999}) is the primary reason why the survival fraction exceeds that predicted by an exponential decay curve (Eq.~\ref{eq:frac_survival}). This is evident in the last panels of Fig.~\ref{fig:hist_U05} and Fig.~\ref{fig:hist_U2}, where the tail of the numerical $f_\text{survive}$ is better described by a power-law decay than an exponential one. We find that resonance sticking generally renders the survival fraction prediction invalid beyond $t/t_S \gtrsim 5$ (see also HT24, figure~10). Within $t/t_S \lesssim 5$, the Fokker--Planck equation provides a reasonable match (usually within a factor of $\sim$$3$) to both the survival fraction and the energy distribution.  

Given the computational intensity of direct $N$-body simulations (e.g., each simulation set required $\gtrsim$5,000 core hours on a modern CPU), the analytical solution (Eq.~\ref{eq:sol_Fokker_Planck_linearized}), which can be computed essentially at no cost, provides predictions accurate to within a factor of $\sim$3 for the scattering timescale $t_S$ and the time evolution of the energy (semimajor axis) distribution, until resonance sticking becomes significant.

\subsection{Dynamical lifetime}\label{subsec:dynamical_lifetime}

The characteristic diffusion timescale in \citet{Duncan.1987} and \citet{Tremaine.1993} (originally denoted as $t_D$ or $t_\text{diff}$, but referred to here as $t_\text{dyn}$ for clarity, as explained below) is defined as the time over which the orbital energy changes by an amount comparable to its own value:
\begin{equation}\label{eq:t_dyn_DQT}
    t_\text{dyn,DQT} \simeq \A_0^{-1/2} \left( \frac{P_p}{M_p^2} \right) \big/ 100,
\end{equation}
where the coefficient $1/100$ was estimated from numerical integrations of highly eccentric, low-inclination, planet-crossing orbits \citep{Fernandez.1981}.  
This characteristic diffusion timescale is \textit{not} the relaxation timescale of the entire distribution (which, by definition, does not depend on the initial $x_0$ or $\A_0$) derived in HT24 and in this work.  
Rather, it represents the ``dynamical lifetime'' of a highly eccentric, low-inclination orbit with semimajor axis $\A_0$, hence the notation $t_\text{dyn}$.

\begin{figure}[htb!]
  \centering
  \includegraphics[width=0.7\columnwidth]{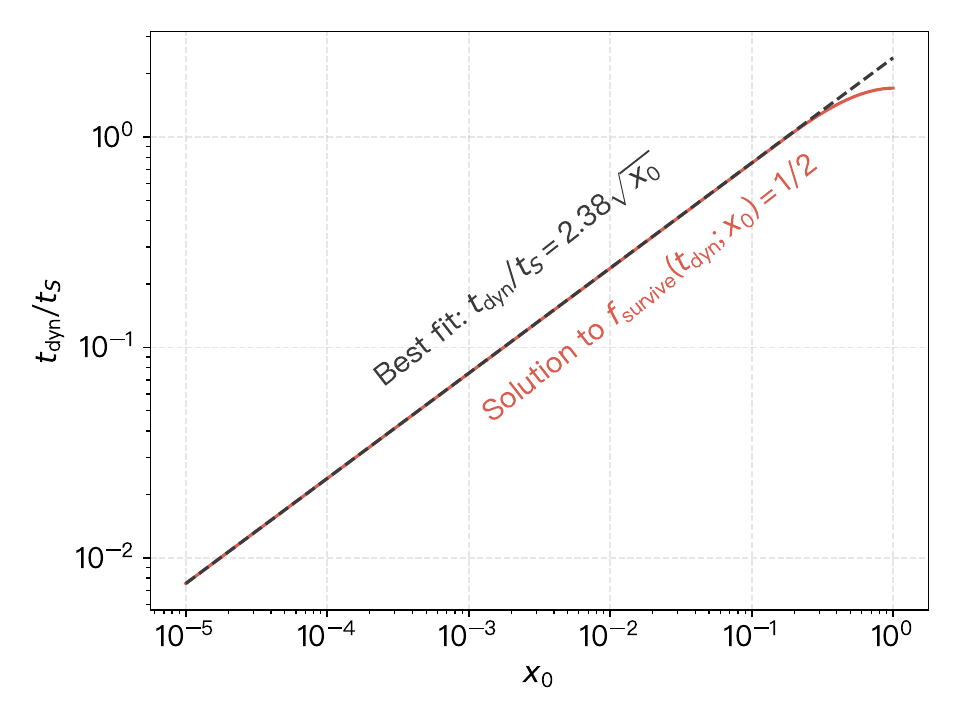}
  \caption{The dynamical lifetime $t_\text{dyn}$ in units of $t_S$ as a function of the particle's initial energy $x_0$. The red solid line represents the solution to Eq.~\eqref{eq:t_dyn_halflife}, and the black dashed line shows the best fit to the red curve in the limit of small $x_0$.}
  \label{fig:t_dyn}
\end{figure}

To directly determine the dynamical lifetime of a particle given initial orbital elements, one can define it as the half-life of a planet-crossing orbit with initial $x_0$:
\begin{equation}\label{eq:t_dyn_halflife}
    f_\text{survive}(t_\text{dyn}; x_0) \equiv 1/2,
\end{equation}
where $f_\text{survive}$ is given by Eq.~\eqref{eq:frac_survival}.  
Although a closed-form expression for $t_\text{dyn}$ is difficult to obtain (since $f_\text{survive}$ is expressed as an infinite series), its value (in units of $t_S$) can be readily estimated as a function of $x_0$, as shown in Fig.~\ref{fig:t_dyn}.

The best-fit solution\footnote{In principle, this relation can also be derived analytically by expanding Eq.~\eqref{eq:t_dyn_halflife} around $x_0 \sim 0$. However, the numerical fit is easier to apply and essentially reveals the same constant, and is therefore adopted.} to Eq.~\eqref{eq:t_dyn_halflife} is given by $t_\text{dyn}/t_S \approx 2.38 \sqrt{x_0}$, which exhibits the same relation $t_\text{dyn} \propto x_0^{1/2} \propto \A_0^{-1/2}$ as in Eq.~\eqref{eq:t_dyn_DQT}.  
To further elucidate the expression for $t_\text{dyn}$, we substitute $x_0 = x_p/\A_0$ to obtain
\begin{equation}\label{eq:t_dyn_halflife_simp}
    t_\text{dyn} \approx 2.38\, \A_0^{-1/2} t_S x_p^{1/2},
\end{equation}
which fully describes the dependence of $t_\text{dyn}$ on the initial $\A_0$ and the Tisserand parameter $\T$, given the full expression of $t_S$ in Eq.~\eqref{eq:relaxation_time_general}. Similarly, by adopting two representative values, $\T = \sqrt{8}$ and $\T = -1$, Eq.~\eqref{eq:t_dyn_halflife_simp} can be rewritten as
\begin{equation}\label{eq:t_dyn_simp}
  \begin{aligned}
    t_\text{dyn,0} &= \A_0^{-1/2} \left( \frac{P_p}{M_p^2} \right) \big/ 270,\\
    t_\text{dyn,retro} &= \A_0^{-1/2} \left( \frac{P_p}{M_p^2} \right) \big/ 3,
  \end{aligned}
\end{equation}
where $t_\text{dyn,0}$ represents the minimum dynamical lifetime (corresponding to the minimum scattering timescale $t_{S_0}$), and $t_\text{dyn,retro}$ represents the dynamical lifetime for retrograde orbits near $\T = -1$.  
It is unclear what exact initial conditions were adopted in \citet{Fernandez.1981} and \citet{Duncan.1987} to numerically estimate the diffusion coefficient, which yielded the constant factor of $1/100$ in $t_\text{dyn,DQT}$ (Eq.~\ref{eq:t_dyn_DQT}).  
However, based on their assumption of highly eccentric, low-inclination, planet-crossing orbits (i.e., $q \sim a_p$ and $i \sim 0$), the initial Tisserand parameter should be approximately $\T \sim \sqrt{8}$ (see Eq.~\ref{eq:tisserand-a-infinity}).  
Therefore, our minimum dynamical lifetime $t_\text{dyn,0}$ is in good agreement with $t_\text{dyn,DQT}$, considering that orbits near $\T \sim \sqrt{8}$  are expected to scatter a factor of a few more slowly than those exactly at $\T = \sqrt{8}$ (the local minimum).

\subsection{Ejection Speed}\label{subsec:ejection_velocity}

Another application of the analytical framework is to estimate the typical ejection speeds of interstellar objects and free-floating planets from scattering. The ejection speed $v_\text{eje}$ is given by 

\begin{equation}\label{eq:v_eje}
  v_\text{eje} = \sqrt{\frac{G M_\star}{|a|}}=v_p \sqrt{\frac{|x|}{x_p}},
\end{equation}
where $a$ is the (negative) hyperbolic semimajor axis of an ejected particle, and $x$ is its normalized energy after ejection. Since we define $0 < x < 1$ for bound orbits, it is thus negative for hyperbolic orbits.

The ejection speed of a scattering particle is determined by the strength of the last energy kick that leads to its ejection. This energy kick is apparently random and determined by the planetary encounter geometry (as detailed in Section~\ref{subsec:averaging}), however, the typical ejection speed can be estimated in the following way:

\begin{equation}\label{eq:v_eje_rms1}
  \begin{aligned}
  \sqrt{\mean{v_\text{eje}^2}}
  &=v_p \sqrt{\frac{\sqrt{\mean{\Delta x^2}}}{x_p}},
  \end{aligned}
\end{equation}
where $\sqrt{\mean{v_\text{eje}^2}}$ is the root-mean-square ejection speed, and $\sqrt{\mean{\Delta x^2}}$ is the root-mean-square energy change of the last kick, which is obtained by setting $x = 0$ in Eq.~\eqref{eq:mean-delta-x2}.

\begin{figure}[htb!]
  \centering
  \includegraphics[width=0.7\columnwidth]{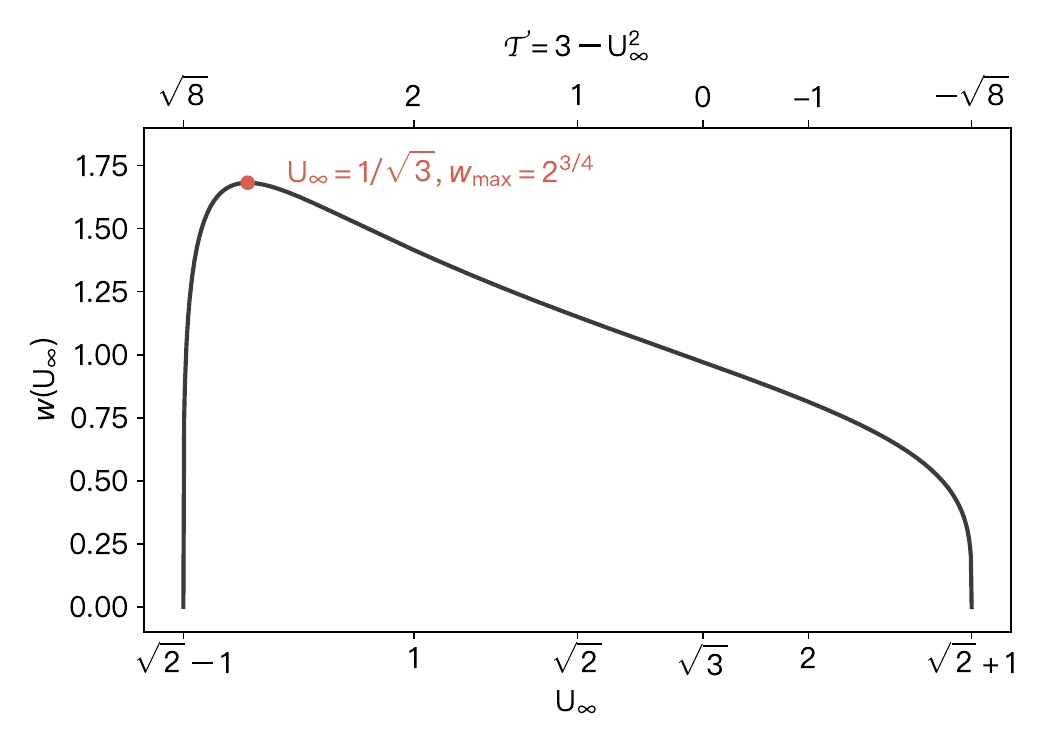}
  \caption{The function $w(\U)$ in Eq.~\eqref{eq:v_eje_rms2} computed from $\sqrt{2}-1$ to $\sqrt{2} + 1$. These boundaries are where an intersecting particle cannot be ejected and one where it would initially be hyperbolic, respectively.
  A maximum $w_{\max} = 2^{3/4}$ is reached at $\U = 1/\sqrt{3}$. The function drops to zero at both boundaries.}
  \label{fig:w_vs_U_inf}
\end{figure}

After simplification, the root-mean-square ejection speed, which we simply call $v_\text{eje}$ from now on, can be expressed as
\begin{equation}\label{eq:v_eje_rms2}
  \begin{aligned}
  v_\text{eje} &= (12\ln \Lambda)^{1/4} \underbrace{\left[ \frac{(\U^2(6-\U^2)-1)^{1/4}}{\U} \right]}_{w(\U)} v_pM_p^{1/3}.
  \end{aligned}
\end{equation}
Similarly, one can approximate the Coulomb logarithm with a constant. We take $\ln \Lambda \approx 5$, and thus the first constant factor in Eq.~\eqref{eq:v_eje_rms2} becomes $\approx$$ 2.8$. The ejection speed's dependence on $\U$ is then fully encapsulated within the second factor $w(\U)$, which is plotted in Fig.~\ref{fig:w_vs_U_inf}. 

One can see that the maximum ejection speed is reached at $\U = 1/\sqrt{3}$, with $w_\text{max} = 2^{3/4} \approx 1.7$. At $\U = \sqrt{2} - 1$, the function $w(\U)$ drops to zero. However, this is not physical but rather indicates the breakdown of the {\"O}pik framework at this boundary. In our framework, a particle can only be ejected if it experiences a close encounter ($B < R_H$) with the planet. This is not the case in real dynamics, where weaker, more distant encounters (i.e., in the diffusion regime) can also eject particles with $\q > 1$. Nevertheless, Eq.~\eqref{eq:v_eje_rms2} provides a useful description of $v_\text{eje}$ as a function of $\U$, which can be further approximated by taking the mean value $\bar{w}(\U) = (\int_{\sqrt{2}-1}^{\sqrt{2}+1} w(\U) \, \mathrm{d}\U)/2 \approx 1.14$ and merging all constants in Eq.~\eqref{eq:v_eje_rms2} into one:
\begin{equation}\label{eq:v_eje_simp}
  \begin{aligned}
  v_\text{eje} &\approx3 v_pM_p^{1/3},
  \end{aligned}
\end{equation}
where $M_p$ is the planet-to-star mass ratio and $v_p$ is the (dimensional) orbital speed of the planet.  This new equation is more useful when the scattering small bodies span a wide range of $\U$ with respect to the planet (such as Oort Cloud comets with an isotropic inclination distribution) or when their initial conditions are uncertain. 

We point out that the ejection speed can also be written as 

\begin{equation}\label{eq:v_eje_vH}
  \begin{aligned}
  v_\text{eje} \approx 4.6 v_H,
  \end{aligned}
\end{equation}
where $v_H = r_H \Omega_p$ is the Hill velocity of the planet, defined as the product of its Hill radius $r_H$ and the Keplerian angular frequency $\Omega_p = 2\pi/P_p$. The Hill velocity is a velocity unit often used in planet formation studies (e.g., \citealt{Goldreich.2004}).

\begin{figure}[htb!]
  \centering
  \includegraphics[width=1.0\columnwidth]{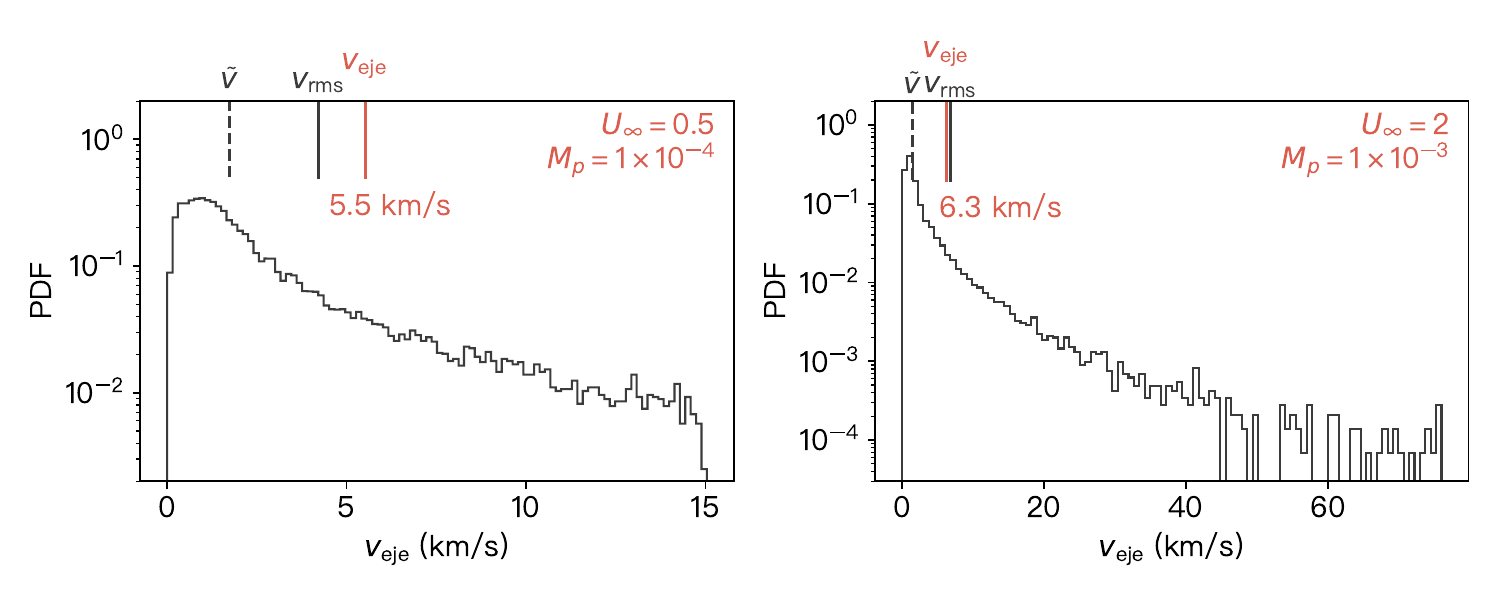}
   \caption{Ejection speed distributions from the two scattering simulations shown in Fig.~\ref{fig:hist_U05} (left) and Fig.~\ref{fig:hist_U2} (right). The black histograms represent the simulated distributions. The theoretical root-mean-square ejection speeds $v_\text{eje}$, computed using Eq.~\eqref{eq:v_eje_rms2}, are shown as red solid lines, while the numerical $v_\text{rms}$ values are shown as black solid lines. Median ejection speeds $\tilde{v}$ are indicated by black dashed lines. In both simulations, the planet is placed on a circular orbit with $a_p = 1$~au, corresponding to an orbital speed of $v_p \approx 30$~km~s$^{-1}$.}
  \label{fig:v_eje}
\end{figure}

To compare our analytical estimate with the numerical results, we plot the ejection speed distributions from the two scattering simulations (see Section~\ref{subsec:verification}) in Fig.~\ref{fig:v_eje}. 
The root-mean-square ejection speed $v_\text{eje}$ predicted by Eq.~\eqref{eq:v_eje_rms2} agrees well with the $v_\text{rms}$ values obtained from the simulations: the first simulation yields $v_\text{rms}= 4.2$~km~s$^{-1}$ (predicted $5.5$~km~s$^{-1}$), while the second gives $v_\text{rms} = 6.9$~km~s$^{-1}$ (predicted $6.3$~km~s$^{-1}$). 
We also note that the ejection speed distribution is \textit{not} a Maxwell-Boltzmann one, instead exhibiting a high-speed tail much more extended than that of a Maxwellian. 
As a result, the root-mean-square speeds are typically a factor of $\sim$3 larger than the median speeds $\tilde{v}$, which are indicated by the vertical dashed lines in Fig.~\ref{fig:v_eje}.

\section{Applications to Planetary Systems}\label{sec:applications}

\subsection{Applications to the Solar System}\label{subsec:applications_solar_system}

\begin{table}
  \centering
  \begin{tabular}{l|cc|ccc|c}
    \hline
    Planet   & $M_p$               & $P_p\ (\mathrm{yr})$ & $t_{S_0}$      & $t_{S,\mathrm{circ}}$ & $t_{S,\mathrm{retro}}$ & $v_\text{eje}$\ (km~s$^{-1}$)  \\
    \hline
    Jupiter  & $9.55\times10^{-4}$ & 11.9              & $51\ \mathrm{kyr}$  & $130\ \mathrm{kyr}$   & $3.0\ \mathrm{Myr}$ & 3.8   \\
    Saturn   & $2.86\times10^{-4}$ & 29.5  & $1.1\ \mathrm{Myr}$ & $3.6\ \mathrm{Myr}$   & $74\ \mathrm{Myr}$  & 1.9 \\
    Uranus   & $4.37\times10^{-5}$ & 84.0              & $110\ \mathrm{Myr}$  & $440\ \mathrm{Myr}$   & $7.7\ \mathrm{Gyr}$ & 0.7\\
    Neptune  & $5.15\times10^{-5}$ & 164.8              & $150\ \mathrm{Myr}$ & $620\ \mathrm{Myr}$   & $11\ \mathrm{Gyr}$ & 0.6\\
    \hline
    \end{tabular}
    \caption{The minimum ($t_{S_0}$), circular ($t_{S, \text{circ}}$), retrograde ($t_{S, \text{retro}}$) scattering timescales, and root-mean-square ejection speed $v_\text{eje}$ for Solar System giant planets. The full expression Eq.~\eqref{eq:relaxation_time_general} is used to compute $t_{S_0}$ and $t_{S, \text{retro}}$ at $\T = \sqrt{8}$ and $-1$, respectively, whereas the approximated Eq.~\eqref{eq:relaxation_time_3} is used to compute $t_{S, \text{circ}}$. Eq.~\eqref{eq:v_eje_simp} is used to computed $v_\text{eje}$.}
  \label{tab:tS-solar-system}
\end{table}

Applying our Eqs.~\eqref{eq:relaxation_time_sqrt8}, \eqref{eq:relaxation_time_3}, and \eqref{eq:v_eje_simp} to the Solar System, one obtains (Table~\ref{tab:tS-solar-system}) 
the scattering timescales and typical ejection speeds for bodies scattered solely by one of the four giant planets. It is not surprising that Jupiter, with its largest mass and shortest orbital period, has the shortest scattering timescale: $\lesssim$1 Myr except for scattering retrograde orbits. The typical ejection speed for Jupiter-crossing objects is $\sim$4~km~s$^{-1}$, consistent with the previous numerical estimate of $\sim$5~km~s$^{-1}$ \citep{Melosh.2003}. 

For Neptune-crossing objects, a scattering timescale of a few hundred million years is consistent with previous numerical simulations (see, e.g., \citealt{Gladman.2006, Huang.2023t}). Moreover, our analytical framework can also be used to quantitatively explain the significant slowdown of scattering for retrograde orbits. For example, retrograde TNOs/Centaurs such as 2008~KV$_{42}$ \citep{Gladman.2009} and 471325~Taowu (2011~KT$_{19}$; \citealt{Chen.2016}) are known to have dynamical lifetimes of a few Gyr. Recent numerical studies on the hypothesized ``polar corridor'' reservoir further show that retrograde TNOs can survive for timescales comparable to the age of the Solar System, despite crossing Neptune's orbit \citep{Namouni.2024}. To the best of our knowledge, there has been no analytical work on high-inclination or retrograde orbits that quantitatively describes the slowdown of the scattering dynamics as a function of $\T$. 
Our Eq.~\eqref{eq:relaxation_time_general} elegantly resolves this fundamental problem in celestial mechanics, showing that Neptune-crossing retrograde TNOs can indeed possess Gyr-scale dynamical lifetimes (Table~\ref{tab:tS-solar-system}).

Estimating the scattering timescales of Centaurs (which cross the orbits of more than one giant planet) and near-Earth asteroids (most of which cross both Earth and Mars before eventually being ejected by Jupiter) is more challenging, as this work is based on a single-planet framework. \citet{Dones.1996} numerically estimated the dynamical lifetimes of six planet-crossing small bodies. For objects with $q \sim 5.6$~au, just outside Jupiter's orbit, their half-lives are $\approx$500~kyr, whereas for objects crossing Saturn's orbit or with $q$ just outside Saturn's orbit, their half-lives range from $1$ to $5$~Myr. These results are generally consistent with the scattering timescales of Saturn and Jupiter listed in Table~\ref{tab:tS-solar-system}, calculated using only the single-planet model.

Additionally, our analytical approach can, in principle, be extended to construct a more general ``scattering network'' applicable to two-planet or even multi-planet systems. 
For instance, to study the dynamical lifetime of near-Earth asteroids, one could first model the scattering dynamics due to Earth analytically, and then account for subsequent scattering by Jupiter once the object becomes Jupiter-crossing. This represents a natural extension of our framework and will be the focus of future work.

\subsection{Applications to exoplanets}\label{subsec:applications_exoplanets}

In the past decade, these has been a burst of detected exoplanets with well-contained planetary orbits and masses, with which a direct computation of their corresponding scattering timescale and small-body ejection speeds is possible with our approach. This is particularly useful when it comes to modeling a planet's interaction with planetesimals in a debris disk (see, e.g.,  \citealt{Lacquement.2025} and \citealt{Lagrange.2025}), as well as modeling the ejection of a planet with a smaller mass by a planet with a significantly larger mass (e.g., the ejection of super Earths by a Jupiter-mass planet, see \citealt{Huangxiuming.2025} and \citealt{Guo.2025}).

\begin{table}
  \centering
  \begin{tabular}{l|cc|c|c}
    \hline
    Exoplanet  & $M_p$               & $P_p$ & $t_{S,\mathrm{circ}}$  &  $v_\text{eje}$\ (km~s$^{-1}$)     \\
    \hline
    Hot Jupiter  & $\sim$$10^{-3}$ & $\sim$$10$ day            & $\sim$$300\ \mathrm{yr}$   & $\sim$30  \\
    Super Earth  & $\sim$$10^{-5}$  & $\sim$$10$ day          & $\sim$$3\ \mathrm{Myr}$  & $\sim$6 \\
    Cold Jupiter   & $\sim$$10^{-3}$ & $\sim$$10$ yr           & $\sim$$100\  \mathrm{kyr}$  & $\sim$4 \\
    Distant Jupiter  & $\sim$$10^{-3}$  & $\sim$$300$ yr        & $\sim$$3\ \mathrm{Myr}$  & $\sim$1\\
    \hline
    \end{tabular}
    \caption{The circular scattering timescales $t_{S, \text{circ}}$ and root-mean-square ejection speed $v_\text{eje}$ for typical exoplanets. }
  \label{tab:tS-exoplanets}
\end{table}

In Table~\ref{tab:tS-exoplanets}, we provide order-of-magnitude estimates of $t_{S,\text{circ}}$ and $v_\text{eje}$ for the four common types of exoplanets. Their rough masses and orbital period estimates are retrieved from \citet{Zhu.2021}'s figures 1 and 2. It is expected that hot Jupiters tend to have the shortest $t_{S,\mathrm{circ}}$ (only $\sim$100~yr) and the highest $v_\text{eje}$ of $\sim$30 km~s$^{-1}$. However, our current method does not yet account for physical collisions with the planet. In reality, a hot Jupiter's Hill radius is only a few times larger than its physical radius; therefore, a significant fraction of objects on crossing orbits will eventually either get tidally disrupted or physically collide with the planet. We plan to incorporate the physical and tidal-disruption radii of both the Sun and the planet into our analytical model in future work, which may be useful for investigating the tidal disruption of Sun-grazing comets \citep{Opik.1966} and the pollution of white dwarfs by exo-asteroids and exo-comets \citep{Veras.2016}.

Cold Jupiters have $t_{S,\mathrm{circ}}$ and $v_\text{eje}$ values comparable to those of Jupiter, whereas distant Jupiters (so far detected primarily through direct imaging) scatter nearby objects on million-year timescales and eject them at slower speeds ($\sim$3~km~s$^{-1}$). 
Super Earths (only known currently near their stars), despite being two orders of magnitude less massive than Jupiter, have scattering timescales comparable to those of distant Jupiters due to their extremely close-in orbits; they also eject small bodies slightly faster than cold or distant Jupiters. It is clear from Table~\ref{tab:tS-exoplanets} that, except for distant Jupiters, hot/cold Jupiters and super-Earths all eject planetesimals or lower-mass free-floating planets with typical ejection speeds comparable to or greater than the velocity dispersion ($1$--$3~\mathrm{km\,s^{-1}}$) of young star clusters \citep{Kuhn.2019}. This may have significant implications for the dynamical evolution of free-floating planetesimals and planets in a cluster environment (see, e.g., \citealt{Wang.2015, Huang.2024j}).

\section{Summary and Discussion}\label{sec:discussion}

We addresses the scattering of an ensemble of particles by a planet using {\"O}pik's close-encounter theory \citep{Opik.1951, Carusi.1990, Valsecchi.2002}, which expresses the heliocentric $(a, e, i)$ of a planet crossing orbit in the $\bm{U}_\infty$ vector, the relative velocity at infinity. We demonstrate that the magnitude of the relative velocity $\U$ (or equivalently the Tisserand parameter $\T$) naturally separates the scattering problem into four regimes, and the {\"O}pik representation is only valid in the closely-coupled regime ($\sqrt{2} - 1 < \U < \sqrt{2} + 1$ or $-\sqrt{8} < \T < \sqrt{8}$).

Based on this representation of the heliocentric orbit, we further average the effect of multiple planetary encounters and obtain the drift and diffusion coefficients for the normalized orbital energy $x$ in Sec.~\ref{sec:fokker-planck}. This enables us to derive the Fokker--Planck equation, which describes the time evolution of the particle number density. With reasonable assumptions, the linearized Fokker--Planck equation can be analytically solved, and a ``scattering timescale'' $t_S$ naturally arises in the solution's exponential decay term. This timescale was only estimated numerically in previous works \citep{Fernandez.1981, Duncan.1987, Malyshkin.1999}, whereas we obtain an explicit expression for $t_S$ as a function of $\T$ (Eq.~\ref{eq:relaxation_time_general}, also see Fig.~\ref{fig:tS_vs_T}). The dynamical lifetime of a particle $t_\text{dyn}$, and the typical ejection speed $v_\text{eje}$ are also derived analytical. 

We highlight below the most convenient forms of our main analytical expressions for the scattering timescale (full expression in Eq.~\ref{eq:relaxation_time_general}), dynamical lifetime (half-life, Eq.~\ref{eq:t_dyn_halflife_simp}), and ejection speed (root-mean-square, Eq.~\ref{eq:v_eje_rms2}).

\begin{enumerate}
    \item \textbf{Prograde, low-$i$ crossing orbits.}  
    For $\T = \sqrt{8}$, scattering occurs most rapidly:
    \begin{equation}
        \begin{aligned}
            t_{S_0} &= \left( \frac{P_p}{M_p^2} \right) \big/ 500,\\
            t_\text{dyn,0} &= \A_0^{-1/2} \left( \frac{P_p}{M_p^2} \right) \big/ 270,
        \end{aligned}
    \end{equation}
    and particles near $\T = \sqrt{8}$ may scatter a factor of a few more slowly than these minimum values.
    
    \item \textbf{Retrograde, high-$i$ crossing orbits.}  
    For $\T \approx -1$, corresponding to retrograde orbits, scattering is the slowest:
    \begin{equation}
        \begin{aligned}
            t_{S,\text{retro}} &= \left( \frac{P_p}{M_p^2} \right) \big/ 7,\\
            t_\text{dyn,retro} &= \A_0^{-1/2} \left( \frac{P_p}{M_p^2} \right) \big/ 3,
        \end{aligned}
    \end{equation}
     and particles spanning a wide range of $\T$ values, from $0$ to $-2$, generally scatter at rates comparable to the retrograde timescale (see Fig.~\ref{fig:tS_vs_T}).
    
    \item \textbf{Initially circular, coplanar orbits.}  
    For $\T = 3$, the particle initially has a low-$e$, low-$i$ orbit near the planet. Although this regime lies beyond the scope of our analytical framework, this work and \citet{Hadden.2024} together constrain the timescale to be:
    \begin{equation}
        \begin{aligned}
            t_{S,\text{circ}} &\approx \left( \frac{P_p}{M_p^2} \right) \big/ 100,\\
            t_\text{dyn,circ} &\approx \A_0^{-1/2} \left( \frac{P_p}{M_p^2} \right) \big/ 50,
        \end{aligned}
    \end{equation}
    
    \item \textbf{Ejection speed for crossing orbits.}  
    Averaging over allowable Tisserand parameters, the typical ejection speed is:
    \begin{equation}
        v_\text{eje} \approx 3 v_p M_p^{1/3}.
    \end{equation}
\end{enumerate}

Scattering of planetesimals by a planet, or smaller planets by a more massive planet, is one of the most fundamental dynamical processes in planetary science. Here we list a few examples where our analytical tool has potential applications:

\begin{enumerate}
  \item \textbf{The formation of the Kuiper Belt, Sednoids, and Oort Cloud.} Neptune's outward migration and scattering of icy planetesimals is thought to play a key role in implanting the hot Kuiper Belt \citep{Nesvorny.2018}, including the sednoid population \citep{Chen.2025}. Conventionally, this problem has been tackled using numerical integrations, which requires a great deal of computational resources. An analytical solution to Neptune's scattering process provides a fast tool for estimating the overall semimajor axis distribution of planetesimals at the early stages of migration, which can then be used as initial conditions to study the early implantation of sednoids (see, e.g., \citealt{Huang.2024s, Hu.2025}). Similarly, the formation of the Oort Cloud is also a process driven by both the galactic environment and planet scattering \citep{Duncan.1987}, where our tool has a potential application.
  \item \textbf{The ejection of interstellar objects and free-floating planets.} Planet scattering is thought to be one of the main mechanisms that ejects interstellar objects and free-floating planets \citep{Portegies-Zwart.2024, Coleman.2025}. Though the exact orbital distribution during the scattering is probably unimportant to this problem, our Eq.~\eqref{eq:frac_survival} provides a good handle of the ejection fraction as a function of time, and Eq.~\eqref{eq:v_eje_rms2} estimates the ejection speed. This can be potentially used to generate more realistic initial conditions for studying the evolution of interstellar objects and free-floating planets in the galactic environment (e.g., the tidal stream of interstellar objects, \citealt{Forbes.2025}).

  \item \textbf{The structure of debris disks and constraining hidden planets.} The interaction between planetesimals and a planet also shapes the structure of a debris disk, and Eq.~\eqref{eq:relaxation_time_sqrt8} provides a decent prediction for the planetary scattering timescale, without requiring any numerical integrations. As an example, we show that the outermost planet of $\beta$ Pictoris ($M_p = 6.4\times10^{-3}$ and $P_p = 23.6$~yr) has a scattering timescale of $t_{S_0} = 1.2$~kyr, consistent with the surrounding region being completely cleared out by itself. However, $\beta$ Pic b alone cannot explain the debris disk's inner edge at $\sim$50~au \citep{Dent.2014}. Suppose the inner edge is open by an undiscovered planet at $a_p = 35$~au, then using Eq.~\eqref{eq:relaxation_time_sqrt8}, we can quickly estimate that to clear debris nearby within the age of the star ($\approx$23~Myr, \citealt{Mamajek.2014}), the planet needs to be at least 0.2~$m_J$ ($M_p = 1.2 \times 10^{-4}$ and $P_p = 155$~yr). Unsurprisingly, this is in excellent agreement with \citet{Lacquement.2025}'s lower bound estimate of 0.15~$m_J$ from numerical experiments.

  As another example, \citet{Lagrange.2025} recently discovered a sub-Jovian planet in the young TWA 7 debris disk using JWST. Using a numerical simulation, they verified that this planet ($M_p = 7\times10^{-4}$ and $P_p = 550$~yr) is able to open a gap at $a = 52$~au in $6$~Myr. We estimate that the scattering timescale of this planet is $t_{S_0} = 2.2$~Myr, comparable to the system age and thus consistent with the idea that the planet is capable of opening a gap in the debris disk. We thus confirm that Eq.~\eqref{eq:relaxation_time_sqrt8} is universal and can be used in any planetary system to estimate the scattering timescale.
\end{enumerate}

\section{Acknowledgments}

We are grateful to Sam Hadden for early discussions on this project; his recent work also motivated us to pursue the full solutions of the scattering dynamics. We also thank Kanata, Konstantin Batygin, Liyong Zhou, Jilin Zhou, Chris Ormel, Xiuming Huang, Qingru Hu, Kangrou Guo, Wenhan Zhou, Alessandro Morbidelli, Cicero Lu, Demetri Veras, and Yasuhiro Hasegawa for many insightful discussions. We thank Giovanni Valsecchi for reviewing the manuscript, which significantly improved its quality. Numerical simulations are conducted on the PC cluster at the Center for Computational Astrophysics, National Astronomical Observatory of Japan. Y.H.~acknowledges support by JSPS KAKENHI grant No. 25K17460. B.G.~is funded by the National Sciences of Engineering Research Council of Canada.  E.K.~acknowledges support by JSPS KAKENHI grant Nos. 24K00698 and 24H00017.

\appendix

\section{Steady state of $\phi$}\label{app:steady-state-phi}

Here, we demonstrate with numerical integrations that regardless of the initial conditions, the angle $\phi$ of an ensemble of scattering particles seems to always relax to a steady state centered around $\phi \approx 45^\circ$.

\begin{figure}[htb!]
  \centering
  \includegraphics[width=0.49\columnwidth]{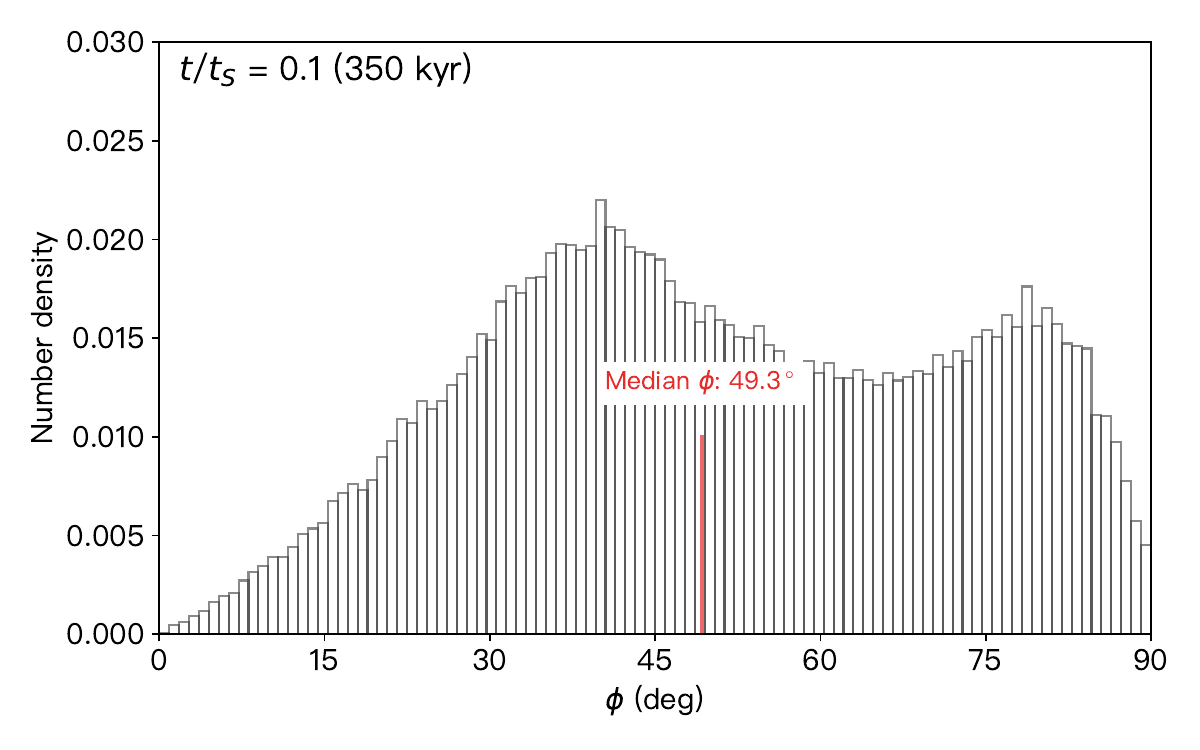}
  \includegraphics[width=0.49\columnwidth]{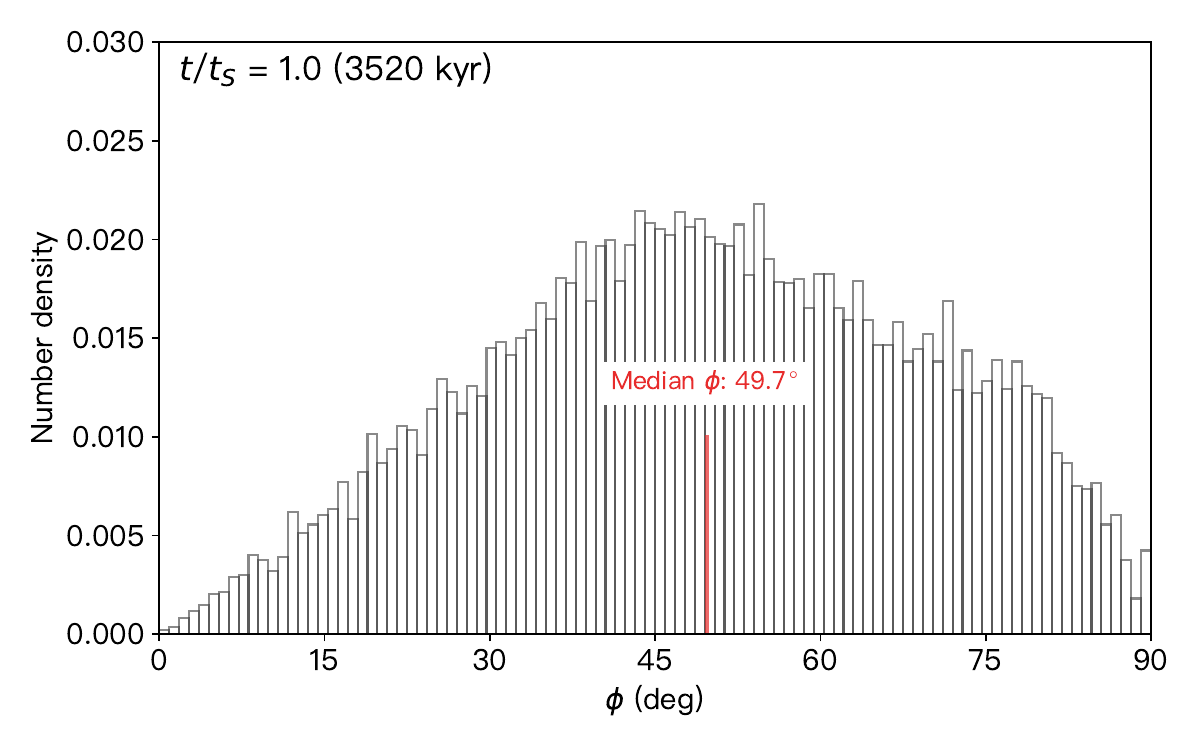}
  \caption{The histograms of $\phi$ for $\U = 1$ ($\T = 2$) particles at $t = 0.1 t_S$ and $1.0 t_S$, respectively. The planet has mass $m_p = 1 \times 10^{-4} M_\odot$ and $a_p = 1$~au, and the corresponding scattering timescale $t_S = 3.5$~Myr is computed using Eq.~\eqref{eq:relaxation_time_general}. The initial conditions for the particles are $\A_0 = 1.1$, $e_0 = 0.854$, and $i_0 = 2.9^\circ$, corresponding to $\phi_0 = 88.2^\circ$. The median $\phi$ during scattering is marked with a red vertical line. }
  \label{fig:phi_sat2}
\end{figure}

\begin{figure}[htb!]
  \centering
  \includegraphics[width=0.49\columnwidth]{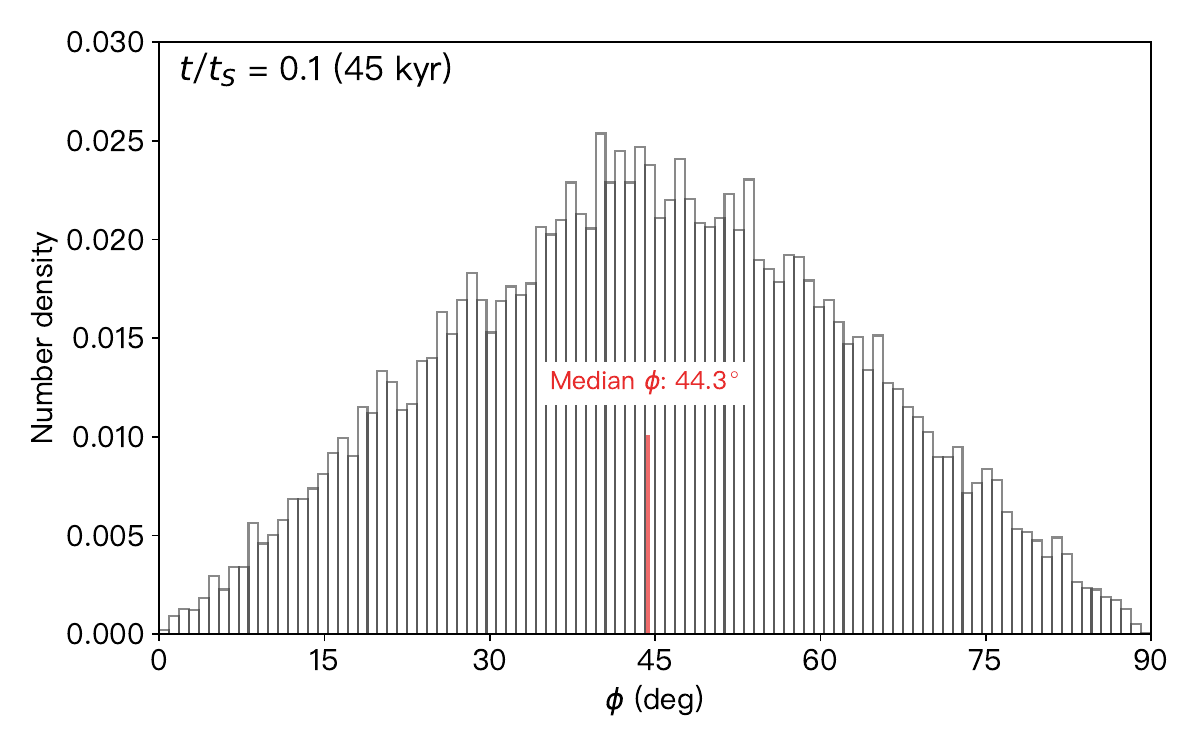}
  \includegraphics[width=0.49\columnwidth]{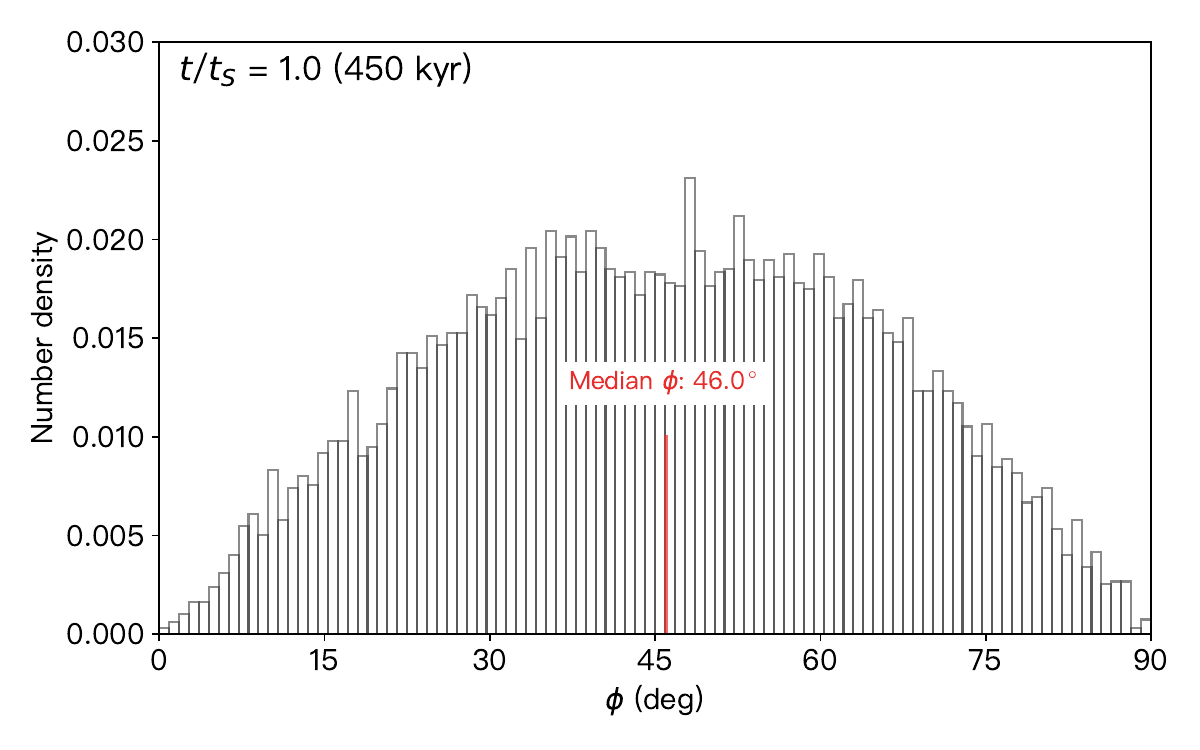}
  \caption{Same with Fig.~\ref{fig:phi_sat2}, but for $\U = 0.5$ ($\T = 2.75$), and $\A_0 = 2.49$, $e_0 = 0.634$, and $i_0 = 15.8^\circ$, corresponding to $\phi_0 = 45^\circ$. The planet has $m_p = 1 \times 10^{-4} M_\odot$, $a_p = 1$~au, and the corresponding $t_S = 450$~kyr.}
  \label{fig:phi_U05}
\end{figure}

As shown in Fig.~\ref{fig:phi_sat2}, these $\U = 1$ ($\T = 2$) particles initially have relatively small inclinations of $i_0 = 2.9^\circ$, which corresponds to $\phi_0 = 88.2^\circ$. The planet has $m_p = 1 \times 10^{-4} M_\odot$ and $a_p = 1$~au, and the corresponding scattering timescale $t_S$ is computed using Eq.~\eqref{eq:relaxation_time_general}. In less than $0.1 t_S$, the median $\phi$ quickly drops to $\approx$$49^\circ$, whereas the distribution of $\phi$ slowly relaxes to a steady state centered around $\phi \approx 45^\circ$ for the rest of the scattering.

Another integration with the same planet but $\U = 0.5$ ($\T = 2.75$) particles is shown in Fig.~\ref{fig:phi_U05}. Their relatively moderate initial inclinations of $15.8^\circ$ correspond to $\phi_0 = 45^\circ$. The initial $\delta$-function distribution quickly relaxed to a distribution centered around $\phi \approx 45^\circ$ and remains largely unchanged for the rest of the scattering.

Therefore, we verify numerically that scattering particles' $\phi$ tend to relax to a steady state centered around $\phi \approx 45^\circ$, which is used as the basis of Eqs.~\eqref{eq:typical-inclination} and ~\eqref{eq:F_function_approx}.

\bibliography{ref_fix_aas}{}
\bibliographystyle{aasjournal}

\end{document}